\pdfoutput=1
\documentclass[12pt]{article}

\usepackage{setspace,graphicx,epstopdf,amsmath,amsfonts,amssymb,amsthm,versionPO}
\usepackage{marginnote,datetime,enumitem,subfigure,rotating,fancyvrb}
\usepackage{natbib}
\usepackage{hyperref,float}
\usepackage{hhline}
\usepackage[toc,page]{appendix}
\usepackage{algorithm}
\usepackage{algpseudocode}

 \graphicspath{ {images/} }

\usepackage{neuralnetwork}
\usepackage{xstring}
\usepackage{tikz}
\usetikzlibrary{arrows,positioning} 
\tikzset{
    punkt/.style={
           rectangle,
           rounded corners,
           draw=black, very thick,
           text width=6.5em,
           minimum height=2em,
           text centered},
}

\usdate

\excludeversion{notes}		
\includeversion{links}          

\iflinks{}{\hypersetup{draft=true}}

\ifnotes{%
\usepackage[margin=1in,paperwidth=10in,right=2.5in]{geometry}%
\usepackage[textwidth=1.4in,shadow,colorinlistoftodos]{todonotes}%
}{%
\usepackage[margin=1in]{geometry}%
\usepackage[disable]{todonotes}%
}

\makeatletter\let\chapter\@undefined\makeatother 

\usepackage{indentfirst} 
\usepackage{jfe}          

\begin{document}

\setlist{noitemsep}  

\title{An Economy of Neural Networks: \\ Learning from Heterogeneous Experiences}

\author{Artem Kuriksha\thanks{I thank Jesus Fernandez-Villaverde and Karun Adusumilli for incredibly valuable discussions and ideas. I also thank Konrad Kording, Annie Liang, and the participants at research seminars at Penn for comments that were very helpful. All mistakes are my own.} \\
  University of Pennsylvania}

\date{}              


\renewcommand{\thefootnote}{\fnsymbol{footnote}}

\singlespacing

\maketitle

\vspace{-.2in}
\begin{abstract}
\noindent This paper proposes a new way to model behavioral agents in dynamic macro-financial environments. Agents are described as neural networks and learn policies from idiosyncratic past experiences. I investigate the feedback between irrationality and past outcomes
in an economy with heterogeneous shocks similar to
\cite{aiyagari1994uninsured}. 
In the model, the rational expectations assumption is seriously violated because learning of a decision rule for savings is unstable.
 Agents who fall into learning traps save either excessively or save nothing, which provides a candidate explanation for several empirical puzzles about  wealth distribution. Neural network agents have a higher average MPC and exhibit excess sensitivity of consumption. Learning can negatively affect intergenerational mobility.
\end{abstract}

\medskip



\thispagestyle{empty}

\clearpage

\onehalfspacing
\setcounter{footnote}{0}
\renewcommand{\thefootnote}{\arabic{footnote}}
\setcounter{page}{1}

\section{Introduction}

This paper aims to achieve three goals. First, I suggest a new model of bounded rationality in macroeconomics.
This model comes from the family of algorithms that successfully learn to perform many important human-like tasks. It also imposes almost no functional restrictions on agents. 
 Second, I attempt to shed light on the problem of learning a decision rule over time, which is different from most of the learning that economics traditionally focuses on. Third, the developed model allows me to investigate the feedback between imperfect decision making and heterogeneous experiences, as policy functions are formed endogenously and depend on agents' past. I focus on the feedback between mistakes in saving and experiences with earnings and wealth. This feedback could have important implications for the economy, but has attracted virtually no attention in the literature.
 
\paragraph{Bounded rationality and heterogeneous experiences.}

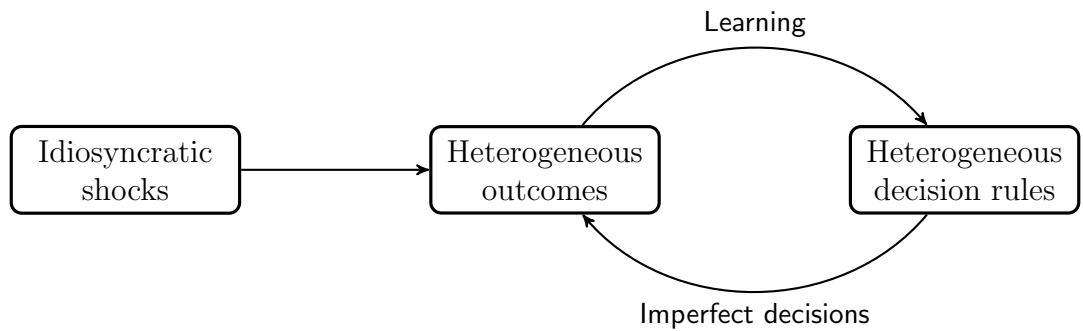
\begin{figure}[!htb]
\begin{center}
\begin{tikzpicture}[node distance=1cm, auto,->,>=stealth',thick]
 \node[punkt, inner sep=5pt] (shocks) {Idiosyncratic shocks};
 \node[punkt, inner sep=5pt, right=2.5cm of shocks] (outcomes) {Heterogeneous outcomes};
 \node[punkt, inner sep=5pt, right=2.5cm of outcomes] (policies) {Heterogeneous decision rules};
 \path[every node/.style={font=\sffamily\small}]
    (shocks) edge node [right] {} (outcomes)
    (outcomes) edge[bend left = 50]  node [right, above, text width=2.0cm, align=center] {Learning} (policies)
    (policies) edge[bend left = 50] node [left, below] {Imperfect decisions} (outcomes);	
\end{tikzpicture}
 \end{center}
   \caption{Feedback between choices and outcomes.} \label{fig:feedback}
 \end{figure}

Figure~\ref{fig:feedback} illustrates the feedback I explore in this study. Consider a model with heterogeneous agents who are subject to idiosyncratic shocks. With rational expectations, only the left arrow is important for the economy. But when perfect rationality is substituted with an experience-based learning process, economic outcomes vary not only because of the shocks but also because of the learned policies. Agents learn to make their choices over time based on what they have observed in the past. Past experiences are heterogeneous across agents and depend both on idiosyncratic shocks and choices made. This creates a feedback between economic outcomes and imperfections in economic choices. In this paper, I suggest a way to model the top right arrow using neural networks, and investigate the feedback that occurs when learning is incorporated into consumption-saving decisions. 

There are two important reasons why idiosyncratic experiences might affect decision making. The first one is formation of expectations. Even if they are perfectly rational in solving their optimization problems, agents with different expectations can make different choices in otherwise identical situations. At the same time, there is substantial evidence that agents with different experiences do have different expectations even about the most crucial and widely discussed economic or financial\footnote{For a psychological reference, see, for example, \cite{hertwig2004decisions}.} variables. 
For example, \cite{nagel2019asset} find that lifetime inflation experiences effectively predict inflation expectations for individuals in the US.
\cite{goldfayn2020expectation} show that East Germans still have higher inflation expectations than West Germans. 
\cite{kuchler2019personal} find that personal experience with employment or the local housing market affects a person's expectations about unemployment or the housing market nationwide. 
\cite{malmendier2011depression} find that those who experienced higher stock returns are more willing to take risks in the future, suggesting formation of expectations as the channel.
\cite{choi2009reinforcement} obtain a similar finding for 401(k) savings. 
\cite{malmendier2016learning} use a model of learning with fading memory to explain patterns in expectations about equity market returns.

The second reason for agents to have different decision rules is that these decision rules are products of a learning process that is endogenous and imperfect. Agents who have different pasts have learned how to act in different environmental circumstances. An agent might be unaware of the right mode of behavior in a situation that is unfamiliar to her. Even if learning was successful for some region of the state space, extrapolation to new regions could be problematic. For example, in the consumption-saving problem, agents that are used to low levels of wealth might fail to save optimally when they suddenly become rich. Similarly, agents that are used to a wealthy life might fail to save correctly when their wealth is suddenly gone. This mechanism also works alongside multiple cognition limitations. For example, agents might have limited memory capacity. They also can overreact to personal experiences or be unable to learn the right way to save in rare events such as unusually high values of productivity. 

The learning process suggested here incorporates both of these reasons. Policy updating is directed toward solving the dynamic decision problem, and thus expectations\footnote{In this paper, the uncertainty is about individual productivity in the future. The model can be generalized to allow agents to choose from a menu of risky assets. In that case, expectations about returns would become important too.} are implicitly taken into consideration. The learning process reflects cognitive limitations and is endogenous and based on past experiences. Thus, the learning artifacts described above are also present.

\paragraph{Learning model.} Each agent faces the income fluctuation problem and has a saving decision rule that is parameterized. Learning takes the form of updating the parameters, which might be different across agents. Specifically, the policy is given by a neural network, and the parameters are its weights. Agents have limited memory of their past. Every period, an agent takes several learning steps. She thinks about what her consumption could have been in the past. The reasoning that she engages in can be informally described as ``I would be happier if I would do this instead of that." Subsequently, the agent makes an incremental update of the parameters.

This model of learning has several advantages. First, it exhibits properties that one naturally expects from a human-like agent in a dynamic environment. At the same time, it is indeed a process of learning that is useful for agents as it is ``consistent." I show that if an agent has favorable hyperparameters (that is, the parameters of the model that are not changed while learning, e.g., the memory size) and learns from experiences that are sufficiently informative about the state space, then the updating rule converges to very good approximations of the optimal policy.

Second, this model is largely inspired by and related to deep reinforcement learning. Reinforcement learning (RL) studies learning guided by rewards and with the goal of choosing optimal actions in an environment.\footnote{See \cite{sutton2018reinforcement} for an introduction to reinforcement learning. See \cite{botvinick2020deep} for a review of recent advancements in deep RL and connections of deep RL with neuroscience.} Incorporating deep neural networks into reinforcement learning (deep RL) has provided a number of breakthroughs in artificial intelligence in recent years. Moreover, neural networks now can achieve a previously unprecedented performance in human-like tasks like image, text, or speech recognition. All that makes neural networks and deep RL a very promising tool for modeling human behavior in economics. 

Third, this model relies on few assumptions about the agents. For instance, I do not rely on many of the ``restrictive assumptions" that are typical in behavioral macroeconomics. I do not assume anything about cognition and attention costs or the structure of incomplete information. Also, I do not assume that agents perform hard cognitive operations. Here, agents do not operate with expectations or other statistical concepts. Neither do they solve any maximization problem or recursive equation. Further, neural networks in principle can approximate almost any function, and thus no restrictive functional assumptions are imposed. Last but not least, the method is generalizable to other dynamic macroeconomic or financial contexts.
See Section~\ref{sec:agents} for the technical description and a detailed discussion of the learning model. 

\paragraph{Findings.} The suggested model allows me to see how agents with potentially different levels of rationality learn to make consumption-saving decisions over time. A major theme of all results I obtain is that learning is a hard task. Learning is prone to instabilities, traps, and biases, which require additional components and assumptions as a remedy. It is a non-trivial task for agents to find a good policy even when they can dynamically experiment with their saving decisions and when they are guided by their previous consumption experiences, which are informative about the stationary Bewley–Huggett–Aiyagari environment the agents live in. This suggests that the assumption of rational expectations is a strong assumption. The model also allows me to estimate the cost of irrationality, which for the main specification I consider is roughly equivalent to one-tenth of consumption.

 The issues that agents face while learning are not specific to this particular task; they also arise in almost all modern applications of RL. This indicates that they can be inherent to the process of learning, and thus it is important to understand their implications for economics. I find that for the consumption-saving problem, they often lead to two particular types of behavior: saving nothing or saving excessively. Agents who are ``lost while learning" in these two ways make the outcomes of the model closer to the data along several margins. They comprise a mass of agents who are persistently stuck around the liquidity constraint (so-called hand-to-mouth status) or they comprise a thicker upper tail in the wealth distribution. This and the fact that heterogeneity in decision making is an additional source of heterogeneity in outcomes increase inequality in the economy. The ones who save nothing also make the average marginal propensity to consume in the economy higher. 
 Another learning artifact I find is inattention to productivity shocks: some agents fail to learn to react to them. This might have implications for economic policy, and in the model it contributes to excess sensitivity of consumption. 
 
The learning process itself also has some interesting properties. Neural network agents make better decisions in familiar regions of the state space, but their biases and mistakes often push them into the regions where their decision rules are severely suboptimal. In particular, descendants of 
parents with little wealth on average know better how to save when they have little savings. However, they often do not know how to save correctly in the ``rich" region of the state space. Surprisingly, that makes them more likely to end up being over-savers. I observe a symmetric pattern for descendants of parents with high wealth, who are more likely to end up with no savings. 

Finally, the noise from mistakes in consumption can decrease the role of wealth inherited from parents. That and the mechanism described above can equalize agents in a succeeding generation. But I also find that in some cases learning can substantially decrease intergenerational wealth mobility. It happens because learning traps might be transmitted through generations. The bulk of the effect comes from the descendants of over-saving irrational agents, who often keep over-saving and generate even more wealth. This is consistent with the data, where large wealth also tends to have high persistency across generations.   
 
\paragraph{Related literature.} 
Methodologically, this paper is a part of the growing number of applications of machine learning in macroeconomics. \cite{duarte2018machine}, \cite{azinovic2019deep}, \cite{fernandez2019financial}, \cite{maliar2019will}, \cite{jfv}, and \cite{kahou2021exploiting} use neural networks to approximate equilibrium objects and solve computationally-hard problems with high-dimensional state and control spaces. Though bounded rationality and learning are not the objects of interest of any of those papers, the idea to update policy parameters using first-order optimality conditions is related to the learning process considered here. 

This paper also utilizes multiple ideas from deep reinforcement learning. The number of applications of reinforcement learning in economics is small but is likely to grow fast. Most of the existing applications are related to allocation of a binary treatment, as in \cite{athey2021policy} or \cite{adusumilli2019dynamically}. This paper is among the very first attempts to use reinforcement learning in macroeconomics or for a continuous economic problem. 
\cite{shi2021learning} also does that and solves the stochastic optimal growth model using deep reinforcement learning. \cite{chen2021deep} solve a model with monetary and fiscal policy and show that a deep reinforcement learning algorithm can recover steady states that are not learnable with adaptive learning. 
The latter two papers also argue that neural networks can be a modeling device for bounded rationality. That said, both keep unexplored many of the economic implications of bounded rationality. The main finding there is that 
RL algorithms can solve 
control problems from macroeconomics. This paper, in contrast, focuses more on the economic implications of imperfect learning and the interpretability and plausibility\footnote{In particular, the approach implemented in this paper allows for learning in a reasonable amount of time. In contrast, in  \cite{shi2021learning}  and \cite{chen2021deep} the agents need hundreds of thousands or even millions of periods to converge to the rational behavior (see Section~\ref{par:ddpg} for a discussion of this issue). In \cite{chen2021deep}, for example, the first phase over which the learning error \textit{increases} takes several hundreds of thousands of periods.  This is problematic if one wants to interpret agents as humans. Further, it might be hard to interpret why real-world agents would have multiple learning episodes. Finally, both papers are concerned with learning by a representative agent. 
At the same time, heterogeneity is a desirable feature for a macroeconomic model of learning and bounded rationality. Learning mistakes are likely to be heterogeneous across agents.} of the assumptions behind learning agents.

Another related project is \cite{zhao2020behavioral}. It also uses neural networks as a device to model behavioral economic agents. However, the focus there is very different. Neural networks are used to model not decision rules but utility, no learning happens, and the macroeconomic implications are not investigated. 

 Conceptually, this project is another attempt to analyze the effects of bounded rationality on the macroeconomy. The vast majority of the literature, however, focuses on the frictions in agents' ability to observe or forecast aggregate variables like the interest rate or inflation. One approach is to postulate the slow spreading of information, as happens in \cite{mankiw2002sticky}. A particularly large class of models follows \cite{sims2003implications}. In these models, learning the true state of the world is costly, and having limited information becomes rational. Deviations from rational expectations can arise then both on the side of firms, as in \cite{mackowiak2009optimal} or \cite{stevens2020coarse}, and on the side of consumers, as in  \cite{gabaix2016behavioral}. Another direction is represented by \cite{farhi2019monetary} and \cite{lowratesgarcia}, who allow for a limited ability of agents to analyze general equilibrium effects while forecasting. 
 
This literature mainly revolves around various questions about monetary policy. The center of attention is agents' ability to process unexpected shocks to inflation and interest rates. After these frictions are considered and the expectations are formed, agents always behave rationally.\footnote{\cite{lian2020mistakes} allows for future mistakes in consumption, but those again are rationally processed in the agents' decisions today.} 
Thus, the possibility of imperfect decision making per se is set aside. Because of that and because the information frictions in these models typically are not experience-dependent, there is no place for a feedback between heterogeneous experiences and bounded rationality. 
This paper, in contrast, abstracts from the potential inability of agents to observe or forecast aggregate variables. Those are constant in the model and agents know them perfectly. The focus is entirely on frictions in decision making and learning decision rules. This learning is experience-dependent, and thus the aforementioned feedback is possible. 


An exception is \cite{ilut2020economic}.\footnote{See also the older version \cite{valchev2017economic}.} Similar to this paper, it is also a model where agents observe the state perfectly but make imperfect decisions, trying to learn a good decision rule over time. 
In \cite{ilut2020economic}, learning happens in a Bayesian way.\footnote{Uncertainty over the policy function is modeled as a Gaussian process, and this generates local spillovers from learning the true policy in a particular point. Reasoning is costly (agents pay for mutual information between the signal and the posterior), and thus an agent might decide not to update her beliefs if her uncertainty at some state is small.} The framework allows the authors to explain facts that the baseline model fails to fit well: high inequality, high marginal propensity to consume, and a large share of agents with zero wealth.
Learning that will be discussed in this paper also allows an explanation of these facts. That said, learning with neural networks has some advantages as a model of bounded rationality. First, the learning model of  \cite{ilut2020economic} is heuristic\footnote{The assumption is that agents can receive signals about the value of the optimal policy function at points of the state space. However, the optimal policy in a dynamic model should solve the recursive problem, in which maximization happens conditional on behaving optimally in the future. This means that it is hard to separate learning in a particular point from learning the policy for other points of the state space. All of that makes Bayesian learning of a policy function less conceptually justifiable than, for example, Bayesian learning of a demand function in \cite{ilut2020paralyzed}.} and omits many details about the important process of ``converting information into a policy function." In that sense, the learning model presented here is more micro-founded. 
Second, the results about the feedback between learning and the past outcomes in \cite{ilut2020economic} come from a specific mechanism \--- agents stick to wrong guesses in states that occur often. This mechanism might  depend on the assumptions about the distributions and the cognition costs. Besides, the role of the choice of the prior distribution is also a concern.\footnote{In particular, it is suggested by the difference in intuition and interpretations between \cite{ilut2020economic} (where the prior coincides with the optimal policy) and \cite{valchev2017economic} (where the prior is constant).}
Finally, the Bayesian approach might not be robust to increases in dimensionality, while real-world agents have high-dimensional state and action spaces.\footnote{For example, because agents are likely to simultaneously solve other economic problems.} I discuss the advantages of neural networks as a modeling device in Section~\ref{sec:agents}.

Finally, this paper is related to the study of systems consisting of multiple agents who adapt and learn according to some algorithm. \cite{holland1991artificial} suggested studying artificial adaptive agents in economics. Back then, the two main tools were classifier systems and genetic algorithms. \cite{arifovic1996behavior} and \cite{bullard1998model} applied this idea in macroeconomics, using genetic algorithm learning in models with overlapping generations. \cite{lettau1999rules} used a learning model with classifiers to suggest a behavioral explanation for excess sensitivity of consumption. In \cite{marimon1990money}, artificial agents learn to use a medium of exchange. The modern research on multi-agent systems is usually interested in agents modeled as neural networks, which have almost completely superseded other RL techniques. See \cite{zhang2019multi} for a review of the literature on multi-agent reinforcement learning. This paper is among the first to analyze the macroeconomy as such a system. Another example is \cite{zheng2020ai}. 

The paper proceeds as follows. Section~\ref{sec:economy} describes an environment in the spirit of \cite{aiyagari1994uninsured}. Section~\ref{sec:agents} describes in detail the implemented model of behavioral agents. Section~\ref{sec:learning} describes the behavior of these agents. Section~\ref{sec:macro} describes the aggregate economic implications. Section~\ref{sec:conc} concludes.

\section{Economic Environment} \label{sec:economy}
The description of the environment is almost identical to the one in \cite{aiyagari1994uninsured}. The two main ingredients of the problem that the agents in the economy have are idiosyncratic income fluctuations and a liquidity constraint. They introduce the precautionary saving motive: agents should save or dissave in order to smooth their consumption, and they do not want to be close to the liquidity constraint because smoothing is impossible there. I will consider two types of agents. Agents of the first type (\textit{rational expectations agents}, or just \textit{RE agents}) will solve this optimization problem in the standard rational way. Agents of the other type (\textit{neural network agents}, or just \textit{NN agents}), which will be introduced in the next section, will learn over time how to save reasonably.
\subsection{Agents}
There is a unit mass of agents, indexed by $i$, who derive utility from consumption $c_{i,t}$ of a numeraire good. Agents discount the future at a rate $\beta$. Productivity shocks introduce fluctuations in an agent's labor income $wz_{i,t}$, where $w$ is the wage in the economy and $z_{i,t}$ is the productivity of the agent. Levels of productivity are independent across agents but are correlated over time for an individual agent. Each agent can smooth her consumption by saving at the interest rate $r$,  but she is subject to a liquidity constraint $\underline{a}.$  Thus, there are two state variables ($a_{i,t}$ and $z_{i,t}$, the wealth and productivity at period $t$), one control variable ($a_{i,t+1}$, the saving for period $t+1$), and each agent maximizes 
\begin{equation}
\mathbb{E}_0 \sum_{t=0}^{\infty} \beta^t u(c_{i,t})
\label{eq:util}
\end{equation}
subject to
\begin{equation}
 c_{i,t} = (1+r)a_{i,t} + wz_{i,t} - a_{i,t+1},\label{eq:budget_constraint}
 \end{equation}
 \begin{equation}
a_{i,t+1} \geqslant \underline{a} .
\label{eq:liq_constraint}
 \end{equation}
 
For the liquidity constraint, I choose the standard conservative option $ \underline{a} = 0$.
The choices for the utility function and the discount parameter are also standard: \begin{equation}
u(c) = \frac{\displaystyle c^{1-\gamma}}{\displaystyle 1 - \gamma}, \ \gamma = 2, \ \beta = 0.96.\end{equation}
An agent's productivity evolves according to a Markov process, so that
\begin{equation}
\mathbb{P}(z_{i, t+1} = z' | z_{i, t}  = z) = \Gamma_{z z'}.
\label{eq:markov}
 \end{equation} The Markov process has 20 states and approximates an AR(1) process given by  \begin{equation} \ln z_{i,t} = \rho\ln z_{i,t-1} + \sigma \epsilon_{i,t},\end{equation} where $\epsilon_{i,t} \sim \mathcal{N}(0,1)$ are i.i.d. The persistence parameter $\rho = 0.6$ and the variance parameter $\sigma = 0.3$ are similar to the values considered in \cite{aiyagari1994uninsured}. 
To obtain the states and the transition matrix $\Gamma$ for the approximating Markov chain, I use the discretization procedure of \cite{tauchen1986finite}. 
  
\subsection{Production}
The production side of the economy is also standard. Output is produced according to
 \begin{equation}
Y_t = K_t^{\alpha}L_t^{1-\alpha}.
\label{eq:prod}
\end{equation}
The capital endowment $K_t$ in the economy is equal to aggregate savings, and the labor endowment $L_t$ comes from the effective labor supply of the agents:  
  \begin{equation} K_t = \int \limits_{0}^1a_{i,t} di, \label{eq:capital} \end{equation} 
    \begin{equation} L_t = \int \limits_{0}^1 z_{i,t} di. \label{eq:labor} \end{equation}
       Because there is an infinite amount of agents, $L_t$ is time-invariant. Capital depreciates at rate $\delta$. The markets for capital and labor are competitive, and thus the interest rate and the wage are respectively
 \begin{equation} r_t = \frac{\partial }{\partial K_t}K_t^{\alpha}L_t^{1-\alpha} - \delta, \label{eq:r} \end{equation}
  \begin{equation} w_t = \frac{\partial }{\partial L_t}K_t^{\alpha}L_t^{1-\alpha}. \label{eq:w} \end{equation}
The values of the productivity parameters are also taken to be standard and equal to $\alpha = 0.33$ and $\delta = 0.1$. There are no aggregate shocks in the economy. 

\subsection{Rational expectations equilibrium}
The benchmark for the analysis is the steady-state equilibrium under rational expectations. I use a standard procedure to find it:
\begin{enumerate}
\item Start with a guess for the equilibrium interest rate $r$;
\item Find the capital $K$ implied by equation~\ref{eq:r} and use equation~\ref{eq:w} to find the wage $w$;
\item Using value function iteration, find the policy $\pi^{RE}(a,z)$ that solves the problem of agents for $r$ and $w$;
\item Find the stationary distribution of savings $a_{i,t}$ for agents following the policy $\pi^{RE}(a,z)$;
\item Find the stationary capital supply $\tilde{K}$ using equation~\ref{eq:capital};
\item Update $r$ to solve $K - \tilde{K} = 0$ (e.g., using the standard binary search algorithm).
\end{enumerate}
The interest rate in the rational expectations equilibrium is 3.29\%. 

\section{Neural Network Agents} \label{sec:agents}

We need a plausible model of human behavior that would allow for bounded rationality in the context of the consumption-saving problem. In the economy described, every period agents should make choices in different points of a continuous state space. Thus, the object they learn is not a finite-dimensional state but a policy function. 
As humans, agents have a limited memory about their past. A learning model should render memories into a decision rule. This can be done with neural networks, which are systems inspired by the structure of human or animal brains. See Appendix~\ref{sec:nnets}  for a brief introduction to neural networks.

\subsection{Learning process}\label{seq:learning_process}
I assume the following model of behavior for neural network agents. At every point, an agent has in mind a stationary policy 
\begin{equation}
\pi(\cdot | \theta): (a_t, z_t) \mapsto a_{t+1},
\label{eq:policy}
\end{equation}
which maps the current state (the wealth and productivity at period $t$) into an element of the action space (the saving for period $t+1$). 
This policy function is given by a neural network with weights $\theta$.\footnote{I clip $\pi(a, z | \theta)$ if the output of the network is below the liquidity constraint. To ensure non-negative consumption, I also clip the output of the network from above if it is too large. I force consumption to be not smaller than the minimal possible labor income. There are two motivations for this. The first one is that no rational agent would ever choose to consume less than $wz_{min}$ at any state. The second is that the ratio of the minimal consumption to the median wage in the model becomes roughly equal to the ratio of the poverty threshold in the US to the median wage in the US.} 
Values of $\theta$ are heterogeneous across agents and determine which decision rules agents follow. Neural networks have high expressivity, and almost any function can be approximated if an appropriate $\theta$ is chosen. 
See Appendix~\ref{sec:nnets} for a brief description of neural networks. See Section~\ref{par:role_nn} for a discussion of why I choose neural networks and not some other function approximator.

The key assumption that defines the learning process is the rule to update $\theta$. As always happens with neural networks and as is likely to happen in human learning, updates are incremental. At every step, the agent sets new weights $\theta'$ to be
\begin{equation} 
\theta' = \theta + \alpha \nabla_{\theta}
\sum_{\substack{\tau \in E}} \big[u(\tilde{c}_{\tau} ) + \beta u(\tilde{c}_{\tau+1} ) 
 \big] 
 = 
\label{eq:learning}
 \end{equation}
 \begin{equation} 
 \theta + \alpha \sum_{\substack{\tau \in E}} \Big( \beta (1+r) u'(\tilde{c}_{\tau+1})  - u'(\tilde{c}_{\tau}) \Big) \nabla_{\theta} \pi(a_{\tau}, z_{\tau} | \theta),
\label{eq:update}
 \end{equation}
 where
 \begin{equation} 
\tilde{c}_{\tau} = (1+r)a_{\tau} + wz_{\tau} - \pi(a_{\tau},z_{\tau} | \theta),
\label{eq:c1}
 \end{equation}
  \begin{equation} 
\tilde{c}_{\tau+1} = (1+r) \pi(a_{\tau},z_{\tau} | \theta) + wz_{t+1} - \tilde{a}_{\tau+2}.
\label{eq:c1}
 \end{equation}
In the above, $\tilde{c}_{\tau}$ and $\tilde{c}_{\tau + 1}$ are counterfactual consumptions that the agent would have if she were to follow the policy $\pi(\cdot | \theta)$. The implied next-period saving $\tilde{a}_{\tau+2}$ is treated as a constant. It is a result of applying a saving policy to the saving choice at $\tau$ that would be made before the learning step:  
 \begin{equation} \tilde{a}_{\tau+2} = \pi\big(\pi(a_{\tau}, z_{\tau} | \theta) , z_{\tau+1} | \tilde{\theta}\big).
\label{eq:next_sav}
 \end{equation}
$E \subset \{t-M, \ldots, t-1 \}$ stands for the set of episodes that the agent uses\footnote{The episodes in $E$ are selected randomly. See Section~\ref{par:memory} for details.} for the learning step.\footnote{I abuse notation here. Counterfactual consumptions $\tilde{c}_{\tau}, \tilde{c}_{\tau + 1}$ are calculated independently for each index $\tau$ of the sum. They are functions of $a_{\tau}, z_{\tau},$ and $z_{\tau+1}$. Equivalently, the learning rule can be defined in terms of triples $(a_{\tau}, z_{\tau}, z_{\tau+1})$ stored in $E$.} Agents are assumed to remember the information about the previous $M$ experiences $\big \{(a_\tau, z_\tau) \big \}_{\tau = t-M}^{t-1}$ they had.
Finally, $\alpha$ is the learning rate, which is the parameter that determines the size of the learning step.

\paragraph{Interpretations.} Before making a new consumption-saving decision, an agent thinks about some episodes that she remembers. As memories are fading and cognitive limitations are present, only a small number of the recent episodes can be processed at once. Thus, she only thinks about a subset $E$ of her previous experiences. She analyzes \textit{ex-post} what saving choices would be better than the ones that the current policy prescribes. She attempts to make an improvement to her utility flow from the episodes she thinks about, which is given by the summation term in equation~\ref{eq:learning}. After that, she makes an adjustment to her policy and uses the updated policy in the future.

Equation~\ref{eq:update}  has the following economic interpretation. The parameters are updated either in the direction of increasing savings or decreasing savings at the states from the memory. The direction of the change is given by a scalar, which is the Euler equation error. 
For example, suppose that an agent remembers some period $\tau$ such that the marginal utility of consumption at $\tau$ under the current policy is much lower than at $\tau+1$. The agent understands that this is a case when she could save more wisely. She is starving right after consuming excessively, and this happens because she is not saving enough. The Euler equation error is positive. Equation~\ref{eq:update} implies that $\theta$ is updated in a way that makes her savings $\pi(a_\tau, z_\tau | \theta)$ larger.

Another interpretation for the updating rule is given by the reinforcement learning literature. \cite{silver2014deterministic} derive an expression for the gradient of the performance objective function with respect to policy parameters. Ideally, agents would use this gradient for all policy updates as it would yield the fastest learning. However, for that they should learn and remember the state-action value function. This seems to be unrealistically hard. Equation~\ref{eq:update} can be considered a feasible behavioral approximation to the optimal policy gradient step.\footnote{The related result is called the \textit{deterministic policy gradient theorem}. The gradient equals the expected value of $\nabla_{\theta} \pi(s | \theta) \nabla_{a} Q^{\pi}(s,a) |_{a=\pi(s | \theta)}$. $Q^{\pi}(s,a)$ is the state-action value, which is the expected reward flow of an agent who is at state $s$, chooses action $a$, and follows policy $\pi$ afterward. Policy gradient algorithms (e.g., DDPG; see Section~\ref{par:ddpg}) usually attempt to estimate $Q$ separately. In practice, this creates another source of instability and requires a very large sample of experiences. However, in the consumption-saving problem this object is more tractable: $Q^{\pi}(s,a) = u(\tilde{c}_{\tau}) + \beta u(\tilde{c}_{\tau+1}) + \ldots,$ where the consumption path is given by the state $s = (a_{\tau} , z_{\tau})$, the choice of action $a = a_{\tau+1}$, and the policy $\pi$. Equation~\ref{eq:update} approximates $\nabla_{a} Q^{\pi}(s,a)$ with $\beta (1+r) u'(\tilde{c}_{\tau+1})  - u'(\tilde{c}_{\tau})$. With this approximation, the agent utilizes her knowledge of her own utility function $u$ and the budget constraint and ignores recurrent derivatives by holding $a_{\tau+2}$ fixed. The latter means that the agent is not considering how a change in her action today would change her actions next period, two periods from now, three periods from now, and so on. This kind of reasoning is likely to be too complicated for a behavioral agent.
}  


\subsection{Discussion}
\paragraph{Role of neural networks.}\label{par:role_nn}
In principle, function approximators of other types (e.g., Chebyshev polynomials) could be used to implement the same logic of learning by updating parameters. Neural networks, however, have advantages that make them a much better choice both for agents and a researcher looking for a credible model of agents' behavior. First, in the real world people usually face high-dimensional problems. Agents not only solve for the consumption-saving trade-off but also divide their portfolios among different assets and choose location, job, working hours, etc.\footnote{Let alone extremely high-dimensional control problems like walking, speaking, and so on and so forth.} Human minds should be able to learn to solve these problems. Neural networks are known to work well in high-dimensional settings for which polynomial approximations seem to be completely infeasible (see~\citealp{bach2017breaking}). Not only are these settings high-dimensional, they are also often related to the problems human minds are solving every day: image, speech, or text recognition, physical movements, etc. It seems reasonable to choose the class of algorithms that works universally for human-like tasks as the device to model human agents. 
Second, neural networks are more stable. For polynomials, there is likely to be a region for which the approximation values are extreme, and it seems unrealistic that a human agent would make decisions like that. Third, neural networks are able to approximate a much wider class of functions 
 (see~\citealp{cybenko1989approximation,barron1993universal}). Fourth, neural networks impose very little prior structure as a modeling device. Thus, the findings do not come from particular assumptions about the functional form of policy functions.\footnote{For example, the agents are not assumed to know that the optimal decision rule is monotone, approximately linear in the points that are far from the liquidity constraint, etc.} Last but not least, neural networks are systems inspired by human or animal brains.

\paragraph{Properties.}
\label{learning_properties}
The described learning process exhibits the following properties:  
\begin{enumerate}
\item \textit{Deviations from rational expectations.} The policy learned by an agent does not have to coincide with the saving policy that maximizes her utility.
\item \textit{Past-dependence.} Agents with different trajectories maximize different functions at every step, and thus end up with different network weights and different policy functions. In particular, different income shocks can make otherwise identical agents learn different decision rules for savings.
\item \textit{Local spillovers.} Neural networks are continuous function approximators. Thus, agents should choose quantitatively similar actions in a couple of states if the two states are close to each other. In the considered environment, the optimal policy is continuous too. Thus, to the extent that an agent was able to learn the policy from the states she observed in the past, she should behave reasonably in states that are not very different from her experiences.\footnote{This property should be amplified by the fact that neural networks usually extrapolate in a linear fashion. The optimal policy in the consumption-saving problem has no non-convexities and can be approximated well locally by a linear function.}
\item \textit{Incremental updating.} When a new experience is processed by an agent, this results in a marginal revision of the policy she had before. As a consequence, even periods that are already forgotten still continue to affect an agent's policy.
\item \textit{Forgetting.} Periods from a long time ago eventually disappear from memory and stop influencing policy updates. Their effect on the policy decays over time, as the agent keeps updating the weights based on newer experiences.
\item \textit{Asymptotic rationality.} All the properties listed above are desirable for a model of bounded rationality. However, one can easily think of a random policy-updating scheme that satisfies these properties but is absolutely meaningless. Such a model of learning would never allow an agent to learn anything useful. What makes the suggested model more credible is that it is \textit{asymptotically rational}. In Appendix~\ref{sec:app_rationality} I show that agents can learn the optimal policy with very high precision in the long run if they have hyperparameters that are sufficiently favorable and have in their memory stack observations that are spread densely over the state space.\footnote{In other words, the universe of economies with NN agents (approximately) nests the rational expectations economy if agents are set to have favorable learning parameters.}
\end{enumerate}
Arguably, all of these properties are shared by human beings. Thus, even if the exact mechanism of the suggested kind of learning might be questioned, it would be a valuable heuristic. It exhibits realistic properties and can be used to investigate the feedback between irrationality and heterogeneous experiences. 


\paragraph{Assumptions.}
A conceptual advantage of this approach is that it assumes that agents use little information. An agent is only assumed to observe her own savings and labor income, the interest rate, and to know her utility function. It is not assumed, for example, that she knows the parameters of the Markov process behind the productivity levels $z_{i,t}$. Another advantage is that an agent is supposed to make only a relatively simple cognitive effort. The agent does not rely directly on any probabilistic calculations, either in Bayesian or in frequentist fashion. She also does not have to solve the Bellman equation or apply the maximum operator.

The policy an agent learns is a result of an ex-post analysis of her life satisfaction. Learning is entirely backward-looking, and the agent essentially is ``preparing to fight the last war." One might doubt that an agent can learn much this way, doing no statistical calculations and not having the Bellman equation in mind. However, in a stationary environment the optimal policy is time-invariant. The past is representative of the future, and optimization over the past yields an object that is good for the future too. In Appendix~\ref{sec:app_rationality}, I show that an agent with the updating rule from equation~\ref{eq:learning} typically converges to the optimal policy if her memories are representative about the state space, her learning parameters are not restrictive, and she has enough time to learn. 


\paragraph{Generalizability and limitations.}
The suggested type of learning is meaningful because the environment in which agents operate is stationary, and repeated attempts to improve counterfactual past experience result in a policy that is good for the future as well. This type of logic might be generalized to other economic choices that are repeated and for which the optimal policy is stable. For example, one could incorporate a menu of risky assets into the model in exactly the same way as was done with the risk-free asset in the model. Alternatively, one can add working hours as a control variable. In that case, the utility flow in equation~\ref{eq:learning} should be augmented with the disutility from labor, and the gradient step in equation~\ref{eq:update} would also include the static labor supply optimality condition. In a similar fashion, neural network agents can in principle substitute rational agents in any DSGE model with tractable optimality conditions. 

At the same time, this constitutes a challenge to generalize this method to some other economic decisions. If an agent faces a non-stationary environment, the optimal behavior in the future might be impossible to learn solely based on the past. Similarly, this type of learning cannot be applied to once-in-a-lifetime decisions like the decision to get higher education. An inherent property of such problems is that an agent cannot learn a good policy by reflecting on her own past. The agent needs to incorporate some external information about the problem (e.g., about rewards following the possible choices) and process this information in a way that is cognitively less trivial than the incremental adjustment of parameters (e.g., by calculating payoff expectations or solving a utility maximization problem).

\paragraph{Differences with other learning approaches.}\label{par:ddpg}
The problem of making choices dynamically and learning a policy that yields a high expected return is at the core of reinforcement learning. The problem of the consumption-saving trade-off is continuous. The bulk of the methods that are popular these days for continuous control problems are model-free actor-critic methods, e.g., DDPG presented in 
\cite{lillicrap2015continuous}. However, these methods are designed for very different learning environments, for example, the ones related to robotics. That implies some conceptual differences, because rewards and transitions between states typically are not known in advance. In particular, the process of learning the critic network usually only uses the information from the observed flow of rewards. For the consumption-saving problem this would mean that agents are unrealistically ignorant about the environment. Essentially, they would have to estimate their own utility function over time, and are not even assumed to know that the saving they choose today will be their wealth tomorrow. This flexibility comes at a cost: the algorithms are very sample-greedy. Just as happens in \cite{shi2021learning} and \cite{chen2021deep}, learning a good policy might take a very large number of periods (well above any feasible number of periods in a human life). In addition to that, it is not particularly realistic that behavioral agents store the state-action value function, which is a very complex object, in their minds.
All of that makes it problematic to use DDPG and other model-free actor-critic algorithms as a modeling device for human behavior in the context I consider here. 

\subsection{Environment}

Though general equilibrium can be calculated in principle, I restrict attention to partial equilibrium. That should be enough to provide the main insights about the differences between rational agents and neural network agents. I hold the wage $w$ and the interest rate $r$ fixed at the equilibrium levels for the rational expectations economy. In other words, I investigate the life of neural network agents in the world of rational agents.

\paragraph{Generations.} Naturally, for rational agents all of the presented statistics are calculated for the stationary distribution. For NN agents, however, the limiting distribution for a particular agent might neglect some important dynamics on the learning path. Given that in the real world agents do not live infinitely,  mistakes specific to the first periods of learning are likely to be more relevant than some ``stationary irrationality" in the model. For this reason, I assume a generational structure about NN agents. 

Specifically, I assume that each agent lives for 100 periods. When an agent is born, she has some initial policy parameters $\theta_{i,0}$. She starts to learn according to equation~\ref{eq:learning} but for the first 20 periods she does not make decisions herself. Instead, her decisions are made by her parent, who follows the rational policy. At age 20, she inherits the wealth and the productivity of the parent, and after that, her savings evolve according to her own decisions $a_{i, t+1} = \pi(a_{i,t}, z_{i,t} | \theta_{i,t} )$. This approach is motivated by two considerations. For one thing, the first 20 periods should serve as a ``burn-in" and decrease the role of the choice for initialization of weights. For another, this allows me to investigate the role of learning in the intergenerational wealth dynamics (see Section~\ref{subsec:mobility}). To calculate aggregate variables, I use the values for adult NN agents only.\footnote{Thus, what I obtain are statistics for the subpopulation of NN agents assuming that they are rarely born among rational agents and there is an equal mass of NN agents of each age. RE agents are either immortal or arrive in altruistic generations.}  

\paragraph{Cap on savings.} Some NN agents will save as much as they can, which results in exponential wealth accumulation. As will be discussed later, this behavior has an interpretation. However, to limit the role of these outliers I restrict possible savings in the economy. The upper limit is set to be roughly twice as large as the maximum observed wealth for RE agents.

\subsection{Other components of the model}

I find that learning is subject to instabilities, which will be discussed in more detail in this and the next section. These instabilities say something important about the process of learning and might have economic implications. They are also universal for deep reinforcement learning. Typically, some additional components are added to an algorithm that are not directly related to the main logic but serve the purpose of improving the convergence properties. Because a model in which agents only have pathological behavior would be unrealistic and provide little insight, I equip NN agents with some modifications too. 

\paragraph{Learning frequency.}
A single gradient step results only in a small update of the network weights. If the initial policy was far from the optimal policy, it would require a large number of steps to approximate the latter well even if the updates are always in the right direction. To make agents plausibly successful in learning, I assume that they take several learning steps per period. 
The interpretation for this is that agents are likely to think about their consumption-saving decisions more than once per period. This is especially realistic in a model where one period stands for a year and the corresponding consumption-saving decision should be a result of many smaller consumption decisions that an agent makes within this period. Another interpretation is that agents think only once, but each ``round of thinking" consists of multiple learning steps. 

\paragraph{Memory structure.} \label{par:memory}
In equation~\ref{eq:learning}, the set of episodes $E$ that an agent uses for learning has the following structure. Each agent has a memory buffer that contains her last $M$ experiences. %
 At each learning step, a random subsample of those experiences is selected.  This technique is called experience replay and helps to mitigate the problem of correlation in the sample of experiences. See \cite{mnih2015human} for a discussion and a biological interpretation of experience replay.  The size of $E$ is called the \textit{batch size} in the literature. Potentially, agents might also use the experiences of other agents for learning updates. 

\paragraph{External experiences.}
I find that stability of learning depends crucially on how informative is the sample of experiences about the state space as a whole.\footnote{This is not surprising. The problem of each agent is dynamic. To make a good saving decision at some point of the state space, an agent should take into consideration what she would choose in the states in which she might be next period. The accuracy of that knowledge in turn depends on the knowledge about other states, and so on. At the end, having a good idea about all points of the state space that might possibly be reached is important for successful learning.} At the same time, the experiences from an agent's past are usually clustered in one particular region of the state space. This is so even for rational agents, as wealth tends to be quite persistent both in the real world and in the Aiyagari economy. For agents on the path of learning, this problem can be exacerbated if their policies are degenerate (e.g., always prescribe no saving) and lead to even less exploration of the state space. I consistently find that these issues can undermine learning with a very high probability.

I assume that each agent considers a fraction of experiences in $E$ that are not her own and instead are generated randomly.\footnote{
To take a learning step, an agent needs a triple $(a_{\tau}, z_{\tau}, z_{\tau+1})$ that describes a level of wealth and a transition of productivity. Thus, I assume that agents have a fraction of randomly generated experiences $(a, z, z')$ in $E$. Values of $z_{}$ are drawn from the stationary distribution of $\Gamma$, and values of $z'$ are drawn conditional on $z_{}$. Values of $a_{}$ are drawn from the uniform distribution over $[0,30]$. The support spans all values of savings that RE agents are likely to reach.} There are two complementary reasons for an agent to learn from experiences that are not her own. The first interpretation is that those are the experiences of her friends\footnote{There is empirical evidence that people pay attention to the experiences of their friends while making economic decisions. For example, \cite{bailey2018economic} present evidence that  this is the case for the decision to buy a house. \cite{bailey2019peer} find that this also happens with the decision to buy a new smartphone.} and other members of society.\footnote{This is also very similar to the idea of learning from experiences generated by several parallel agents, which is exploited by asynchronous reinforcement learning algorithms. See, for example, \cite{mnih2016asynchronous}.}$^{,}$\footnote{This is also similar to exploration \--- adding random noise to actions, which is implemented in many RL algorithms. However, the goal of exploration there is to obtain new information about the environment. The standard exploitation-exploration trade-off is not applicable in the context of this paper as agents know the environment already. However, to make learning more stable, they need to think about the states that are not from their own experience.}
When an agent is informed about the income and wealth of a friend, she starts to think what she would do in his shoes. Similarly, seeing a poor or rich person on the street makes her think about what she would do if she were poorer or richer. The second interpretation is that those are hypothetical experiences that an agent considers for the sake of better learning. These thought experiments are needed to learn successfully
 and it is natural to assume that they are a part of the human reasoning process. 

\paragraph{Initialization.}
At the beginning, agents start with policies that are close to the ``inertial" decision rule. That is, each agent starts with $\theta_0$ such that
$\pi(a, z | \theta_0)  \approx a$.\footnote{This is achieved by setting $\pi(a, z | \theta) = a + \mu \phi(a, z |\theta),$ where $\phi(a, z | \theta)$ is the output of a neural network. Here $\mu = 0.01$ is a small number, and $\mathbb{E}_{\theta_0} \Big[ \phi(a, z | {\theta_0}) \Big] = 0$ as weights ${\theta_0}$  are initialized randomly with the standard procedure of PyTorch. Thus, at first all agents have the inertial decision rule with some little idiosyncratic noise. See \cite{NEURIPS2019_9015} and PyTorch documentation for the details about the random initialization of weights.} 
An agent following this rule does not exercise any cognitive effort and just copies her past savings. For example, if the agent has some positive wealth, then she simply keeps her amount of savings the same and consumes all her interest income together with her labor income. Such a passive approach to savings seems to be the most realistic candidate for a starting point of learning. It is also a rare example of a simple and natural decision rule that is correctly defined at any state and for any feasible liquidity constraint $\underline{a}.$\footnote{Saving $a$ at a state $(a, z)$ is possible when consumption $a(1+r) + wz - a$ is positive. At the same time, to insure that consumption is always positive, it should be that  $
\underline{a}(1+r) + w z_{min}  - \underline{a}> 0.$ Thus, the first inequality follows from $a \geqslant \underline{a}$. See \cite{aiyagari1994uninsured} for more details.}

Besides its behavioral interpretation, this initialization has the advantage of introducing from little to no bias. First, the optimal policy in the Aiyagari economy is quantitatively close to the diagonal line. Agents save slightly above their previous saving for small values of wealth and  slightly below it for large values of wealth. Second, because of the chosen generational structure, NN agents start with the same distribution of wealth as RE agents. Agents who always copy their savings from the past would keep this distribution unchanged. Thus, if we observe some difference between the wealth distribution of NN agents and the distribution of RE agents, this difference comes from something that NN agents learned and not something they started with.

\paragraph{Target network.} \label{par:target}
For the next-period saving defined in equation~\ref{eq:next_sav}, $\tilde{\theta}_t$ does not coincide with $\theta_t$. Instead, it slowly tracks the policy network with the updating rule given by $\tilde{\theta}' = (1-\lambda)\tilde{\theta} + \lambda \theta$, just as happens in \cite{lillicrap2015continuous}.
In reinforcement learning, the network with the second set of weights $\tilde{\theta}$ is usually called a target network. In practice, target networks are essential for stability of learning. The intuition behind the need for a target network here is similar to the one discussed in \cite{mnih2015human}. Consider the following example. As can be seen from equation~\ref{eq:update}, a larger value of $\tilde{a}_{\tau+2} = \pi(a_{\tau+1}, z_{\tau+1}) | \tilde{\theta})$ implies smaller $\tilde{c}_{\tau+1}$, and thus an update of $\theta$ toward larger savings. If $\tilde{\theta}_t$ is set to be the same as $\theta_t$, then an over-saving bias in the policy leads to an update that is also biased toward over-saving. Intuitively, if an agent is certain that she must save a lot for two periods later, then the logic of consumption smoothing requires her to save a lot for tomorrow. Thus, a biased policy can become even more biased after the learning step. The agent might be trapped in an ``over-saving chase." Cycles of this kind can lead to divergence or oscillations. A slowly updated target network dramatically mitigates this problem as the implied next-period savings become more stable.

One possible interpretation of this aspect of learning is that it is something like an evolutionary-developed mechanism (which all AI algorithms have to utilize too). Agents are reasonably cautious of the described ``learning oscillations" and use a conservative value for the next-period saving in the learning rule given by equation~\ref{eq:learning}. Another interpretation is that agents' cognitive costs make it too hard to simultaneously optimize the choice at some state and to consider the effects of the previous policy updates on the next-period saving. In the extreme case, agents would just optimize over $\pi(a_{t}, z_{t} | \theta),$ holding the next-period saving at the same level actually chosen in the past, that is, setting  $\tilde{a}_{\tau+2} = a_{\tau+2} = \pi(a_{\tau+1}, z_{\tau+1} | \theta_{\tau+1})$. However, that would be extremely inefficient in terms of learning.\footnote{The value of $\pi(a_{\tau+1}, z_{\tau+1} | \theta)$ that solves the intertemporal trade-off is close to the rational choice at $(a_{\tau+1}, z_{\tau+1})$ only if $\tilde{a}_{\tau+2}$ is close to the rational choice. In contrast, if the value of $\tilde{a}_{\tau+2}$ is a suboptimal saving choice, it biases the learning update. Thus, if $\tilde{a}_{\tau+2}$ is not updated and always equals $a_{\tau+2}$ chosen with some imperfect policy, then this experience biases learning until it is forgotten.} A target network can be thought of as adjusting for the fact that the policy was updated but with some delay.

\paragraph{Optimization.} Instead of making simple gradient steps of a fixed size, the optimization from equation~\ref{eq:learning} is done with Adam. Adam is a gradient-based optimization algorithm introduced in \cite{kingma2014adam}. It is designed for training neural networks and is essentially the standard in modern deep learning. It facilitates stability and increases the speed of convergence.

\subsection{Algorithm} Algorithm~\ref{alg} shows all elements of the model together. In the pseudocode, $T$ stands for the length of life, which is 100 for all agents in the model. The age when childhood ends is 20 for all agents. $F$ stands for the learning frequency. The memory buffer $R$ keeps the last several personal experiences of the agent and has a finite size $M$. The batch size is $N = | E |,$ where $E$ is the set of episodes used for a learning step. The share of external experiences is $\kappa$. The values of these hyperparameters are discussed in Section~\ref{sec:cogn_margions}. The cap on savings $a_{max}$ is 50, the minimum consumption $c_{min}$ equals $wz_{min}$ (see the discussion in Section~\ref{seq:learning_process}). 

\begin{algorithm}
\begin{algorithmic}
\caption{Algorithm of neural network agents.}\label{alg}
\State Initialize policy network $\phi(a,z | \theta)$ and target network  $\phi(a,z | \tilde{\theta})$
\Function{$\pi(a, z | \theta)$}{}
     \State $x =  a + \mu \phi(a, z |\theta)$ 
          \If{$x < \underline{a}$}
      \State $x = \underline{a}$
    \EndIf
     \If{$x > a_{max}$}
      \State $x = a_{max}$
    \EndIf
     \If{$ (1+r)a + w z - x < c_{min}$}
      \State $x = (1+r)a + w z - c_{min}$
    \EndIf
    \State\Return $x$
\EndFunction
\State Initialize memory buffer $R$ of size $M$
\State Draw initial saving $a_1$ from the stationary distribution of wealth for RE agents
\For{$t = 1, \ldots, T$}            
	\State Draw $z_t$ from Markov process $\Gamma$
	\If{$t \leqslant $ (end of childhood age)}
		\State Select action $a_{t+1} = \pi^{RE}(a_t, z_t)$
	\Else 
		\State Select action $a_{t+1} = \pi(a_t, z_t | \theta)$
	\EndIf
	\State Store episode $(a_{t-1}, z_{t-1}, z_{t})$ in R 
	
 	\For{step $ = 1, \ldots, F$}     
		\State Reset learning buffer $E$ 
		\State Draw the number of external experiences $k$ from the binomial distribution $B(N,\kappa)$
		\State Draw $k$ artificial experiences $(a, z, z')$, put them into $E$
		\State Sample $N-k$ random experiences from $R$, put them into $E$
		\State For each episode $e = (a, z, z') \in E$, set $$\tilde{c}_{e,1} = (1+r)a + wz - \pi(a_{},z_{} | \theta),$$
$$\tilde{c}_{e,2} = (1+r) \pi(a_{},z | \theta) + wz' - \tilde{a},$$
\State where $\tilde{a}_{} = \pi\big(z', \pi(a_{}, z_{} | \theta)  | \tilde{\theta}\big)$ is treated as a constant.
		\State Update weights $\theta$ by maximizing $$\sum_{\substack{e \in E}} \big[u(\tilde{c}_{e,1} ) + \beta u(\tilde{c}_{e,2} ) \big] $$
	 	\State Update target network: $$\tilde{\theta}' \gets (1-\lambda)\tilde{\theta} + \lambda \theta$$
	\EndFor       
\EndFor
\end{algorithmic}
\end{algorithm}

\section{Learning Outcomes} \label{sec:learning}
One of the broad conclusions of the paper is that learning a good saving policy dynamically from previous experiences is a hard task. There are several obstacles that make the process of learning intrinsically unstable. These learning issues seem to be inherent to learning as many of them universally arise in most of the modern applications of deep RL. Thus, these or similar issues might affect learning and outcomes of real-world economic agents. In the consumption-saving problem, they might lead to biases or learning traps in which agents save nothing or save excessively. All of that suggests that the assumption of rational expectations in macroeconomics is, in fact, a strong assumption.

\subsection{Learning obstacles}
The following are the problems that I find negatively affect learning done by neural network agents. Some of them can be mitigated by the modifications described in the previous section. 
\begin{enumerate}
\item Learning a good policy function requires approximation of a point in an infinitely dimensional space. This implies that in general a good approximation can be achieved only through calibration of a large number of parameters. That in turn requires a lot of updating steps and a lot of experiences to learn from. Thus, a high learning frequency and/or a long learning time are required for successful learning. 
\item The experiences from which agents learn are correlated over time. As agents typically do not accumulate or decrease savings sharply, levels of $a_t$ usually are concentrated in one region of the state space. An agent can mitigate this problem by using external experiences and randomizing experiences from her memory. 
\item The experiences depend on the policies that agents learned. The wealth that an agent has at some period depends on the saving decisions she made before. If the current policy is degenerate (e.g., implies always saving nothing), then the sample to learn from also becomes degenerate (e.g., it consists only of experiences with zero wealth). Again, an agent can mitigate this problem by using external experiences. 
\item The policy learned before an update affects the update. This opens room for self-perpetuating biases, like the over-saving chase described in Section~\ref{par:target}. Target networks help to mitigate this problem. 
\item Learning cannot happen on the boundaries of the action space. If $\pi(a, z | \theta)$ is too small or too large, it must be clipped to be a valid action. This can happen, for example, if an agent always wants to consume more than her cash-on-hand. The gradient with respect to $\theta$ is zero, and there is no update of the weights. If the policy $\pi( \cdot  | \theta)$ is very pathological and this happens at every considered state, the agent is stuck without learning. Though this particular mechanism is somewhat specific to neural networks, it also reflects something about learning in general. Successful learning requires an informative feedback about the quality of the decisions made. When an agent has no chance to meaningfully experiment because the choices she makes are infeasible or too suboptimal, this might result in a learning trap. It is either very hard or impossible to learn if the current policy is degenerate.
\item Learning is slow for high values of consumption. In equation~\ref{eq:update}, marginal utilities $u'(\tilde{c}_\tau)$ and $u'(\tilde{c}_{\tau+1})$ and consequently the Euler equation error become small if $\tilde{c}_\tau$ and $\tilde{c}_{\tau+1}$ are both large. Then, even if the policy update happens in the right direction, it affects the policy very little. This happens if an agent has accumulated a lot of wealth and consumes from the high interest income she has. Intuitively, for wealthy high consumers, consumption mistakes are less noticeable and make them learn less. Though it makes perfect sense for agents to pay more attention to situations when consumption mistakes affect utility a lot, this mechanism can result in persistence of over-saving behavior. \end{enumerate}

Before I present the simulation results, I will discuss how these obstacles become relevant through the choice of the hyperparameters that make agents boundedly rational.   

\subsection{Margins of rationality}\label{sec:cogn_margions}
Algorithms based on neural networks almost unavoidably require a relatively large set of hyperparameters to define. Some of those have a natural interpretation; others are less interpretable. At the end, I use the following research strategy. I choose hyperparameters in a way that provides good learning results. Then, I choose some parameters of the model that I experimentally find important for performance and that are interpretable as ``margins of rationality." I investigate the effect of making agents less rational on these margins.  

Deviations from rationality might take different forms. Very generally, agents might have limited ability to collect, store, or process information. The first happens in the model, e.g., because an agent learns from a sample of correlated experiences. The second happens if she only can remember a few periods from her past. The third happens if the agent is limited in her ability to convert memories into a decision rule, e.g., because few learning steps are taken each period. As there are many reasons for irrationality, there are many hyperparameters that positively affect learning. 
I found the following ones to be important.
\begin{enumerate}
\item \textbf{Learning frequency}. The higher the number of learning steps per period,  the faster agents develop policies that are close to the optimal saving rule. 
\item  \textbf{Share of external experiences}. The higher the share of random external experiences, the fewer the agents who converge to degenerate policies. As was discussed in the previous section, awareness about outcomes of others is important for successful learning.\footnote{At the same time, larger awareness might in principle be harmful as it restricts agents from focusing on the experiences that are more relevant for them. For example, an agent with low wealth might want to prioritize updating policy in the region of the state space with low savings.}


\item \textbf{Network architecture}. I find that if the neural architecture is too simple, then the model is not expressive enough and it is impossible to approximate well the optimal policy.\footnote{However, there is no monotone dependence, as the standard bias-variance trade-off logic applies. A neural network with a complicated architecture and many neurons introduces higher variance.}
\item \textbf{Memory size.} Memory size positively affects learning results, but the effect is moderate. This is because some information from the experiences that are already dropped from memory is still kept in the weights of the network. 
\end{enumerate}

Since all of these hyperparameters are sensible margins of (ir)rationality, I do not choose a single one. To investigate the role of bounded rationality and provide some related comparative statics, I consider two types of NN agents. The second type (\textit{high rationality}) is supposed to strictly dominate the first type (\textit{low rationality}).
The parameters for the two types are presented in Table~\ref{table:conf}. The default option for plots is the first configuration, as it provides stronger insights about the role of bounded rationality.

\begin{table}[h]
\begin{center}
\begin{tabular}{lccc}
\hhline{~===}
 & Low rationality  & High rationality \\ \hline
Neurons in hidden layers           &       (4,4)                   &          (8,8)        \\
Learning frequency  ($F$)          &             50            &        200          \\
Share of external experiences ($\kappa$)         &          25\%            &          50\%        \\
Memory size ($M$)          &                    50            &          100        \\
Batch size ($N$)  &      \multicolumn{2}{c}{10}        \\
$\alpha$ &       \multicolumn{2}{c}{0.01}        \\ 
$\lambda$ &       \multicolumn{2}{c}{0.01}        \\  \hline \hline
\end{tabular}
\end{center}
\caption{Configurations of NN agents.}
\label{table:conf}
\end{table}

The other hyperparameters are less interpretable as margins of rationality and are taken to be the same for both agents. The batch size is 10, which means that at each learning step, agents think about the 10 periods selected in $E$. This number seems to be intuitively plausible. At the same time, I find that batch size affects the results very little. The value of the learning rate is a common choice in applications of reinforcement learning to simple control problems (other common choices would be $10^{-3}$ and $10^{-4}$). I find this learning rate to be a good choice for the trade-off between quality and speed of learning. A learning rate that is too high would result in the standard gradient descent problems like ``overshooting." A learning rate that is too small would make learning very slow (that could be somewhat compensated by increasing the learning frequency, but a very high learning frequency does not seem plausible for human agents). The weight $\lambda$ for updating the target network is also among the standard choices in the literature. A smaller value of $\lambda$ would slow down learning, while a larger value would decrease stability. The current choice implies that NN agents with low rationality almost entirely forget their policy in 3-5 periods (for example, $0.99^{5\times50} \approx 0.08$), which seems to be realistic. The main logic behind these choices is that they can provide a good approximation to rationality. I keep them fixed and focus on the effect of making agents less rational on the dimensions with clear interpretation.  

\subsection{Examples of learning}\label{subsec:examples}
\begin{figure}[t]
    \begin{center}
    \subfigure[Diversion from RE wealth trajectories. Shaded area has width equal to $2 \times$(standard deviation of diversion).]{\includegraphics[width=0.47\textwidth]{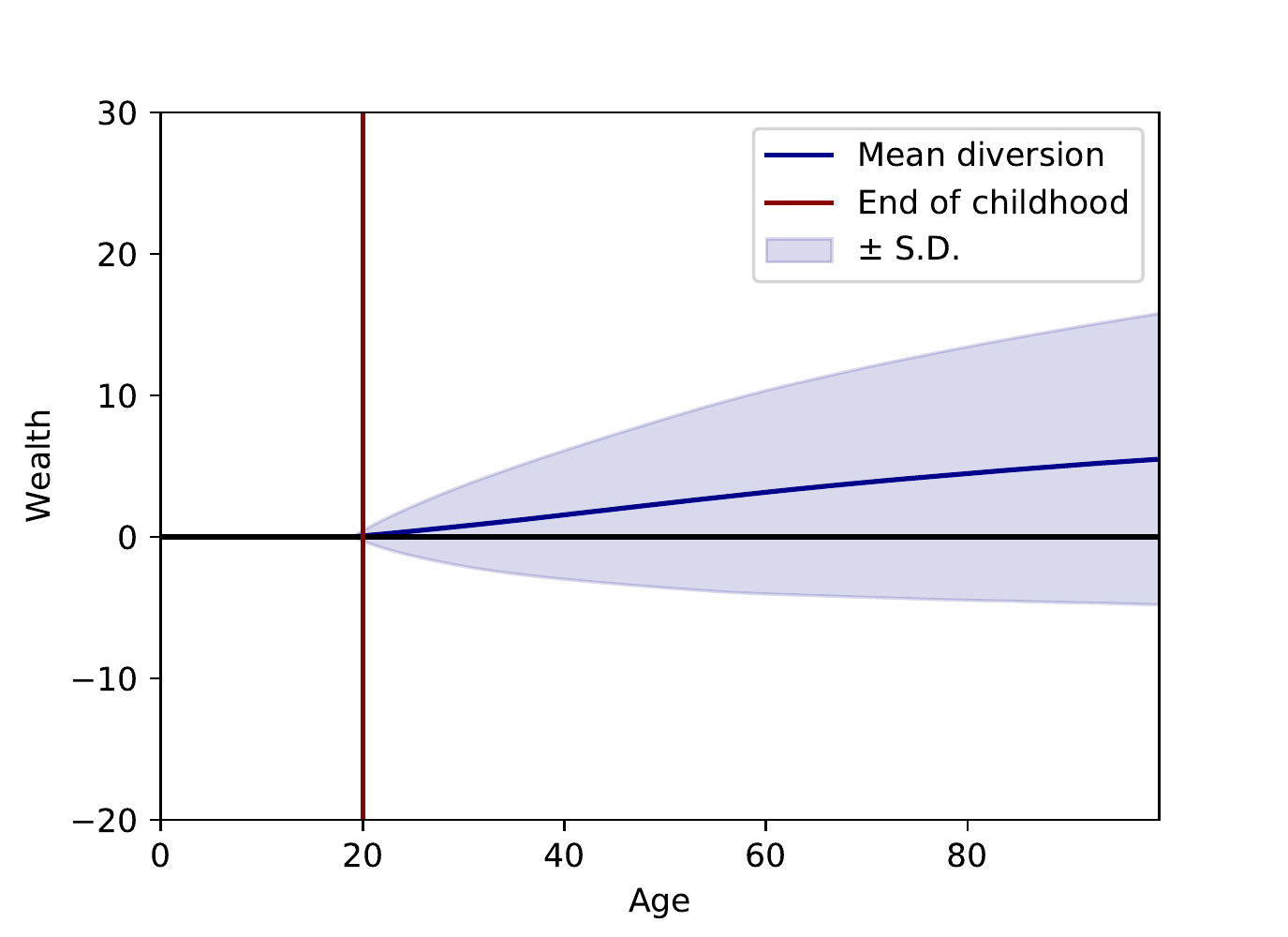} \label{fig:timing}} 
    \subfigure[Wealth trajectories, example.]{\includegraphics[width=0.47\textwidth]{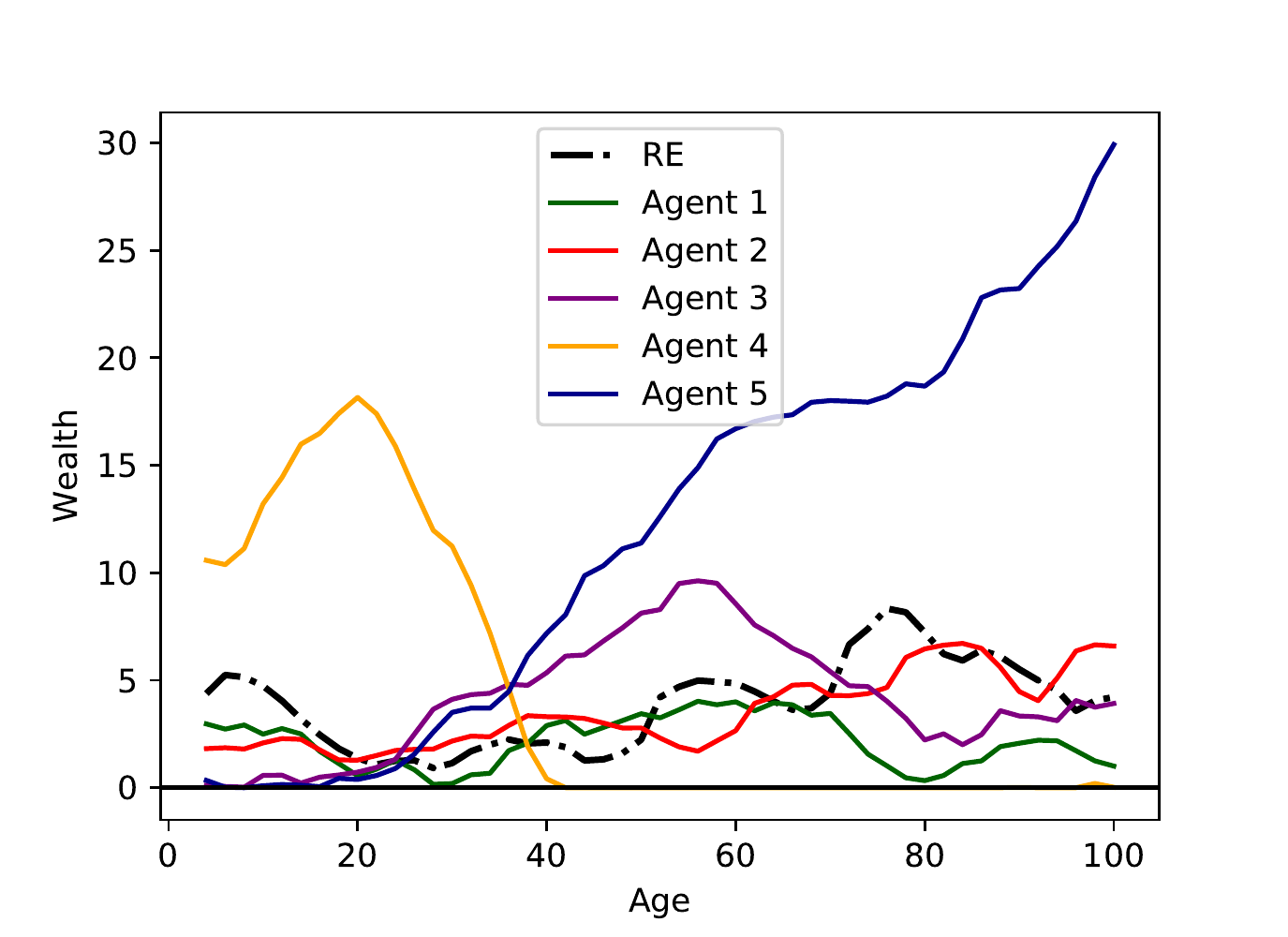} \label{fig:traject}}
    \subfigure[Histories of learning errors, example. \newline Euler error is defined as $\frac{\beta (1+r) u'(c_{t+1})}{u'(c_{t})} - 1$.]{\includegraphics[width=0.47\textwidth]{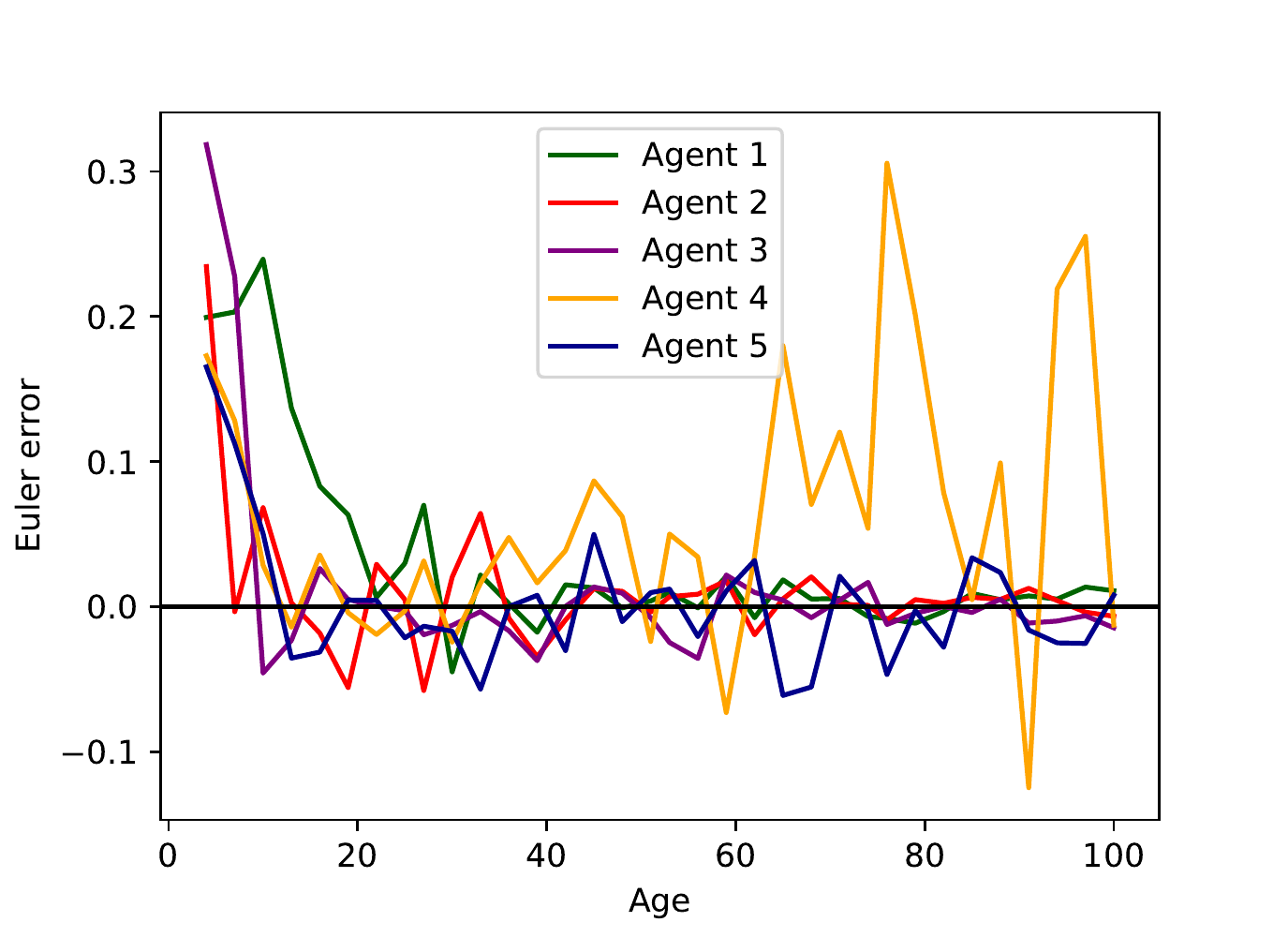} \label{fig:loss}}
    \subfigure[Policies in the last period, example. Saving rate is defined as $\frac{a_{t+1}-a_t}{ra_t + wz_t}$, $z$ fixed at the mean level. ]{\includegraphics[width=0.47\textwidth]{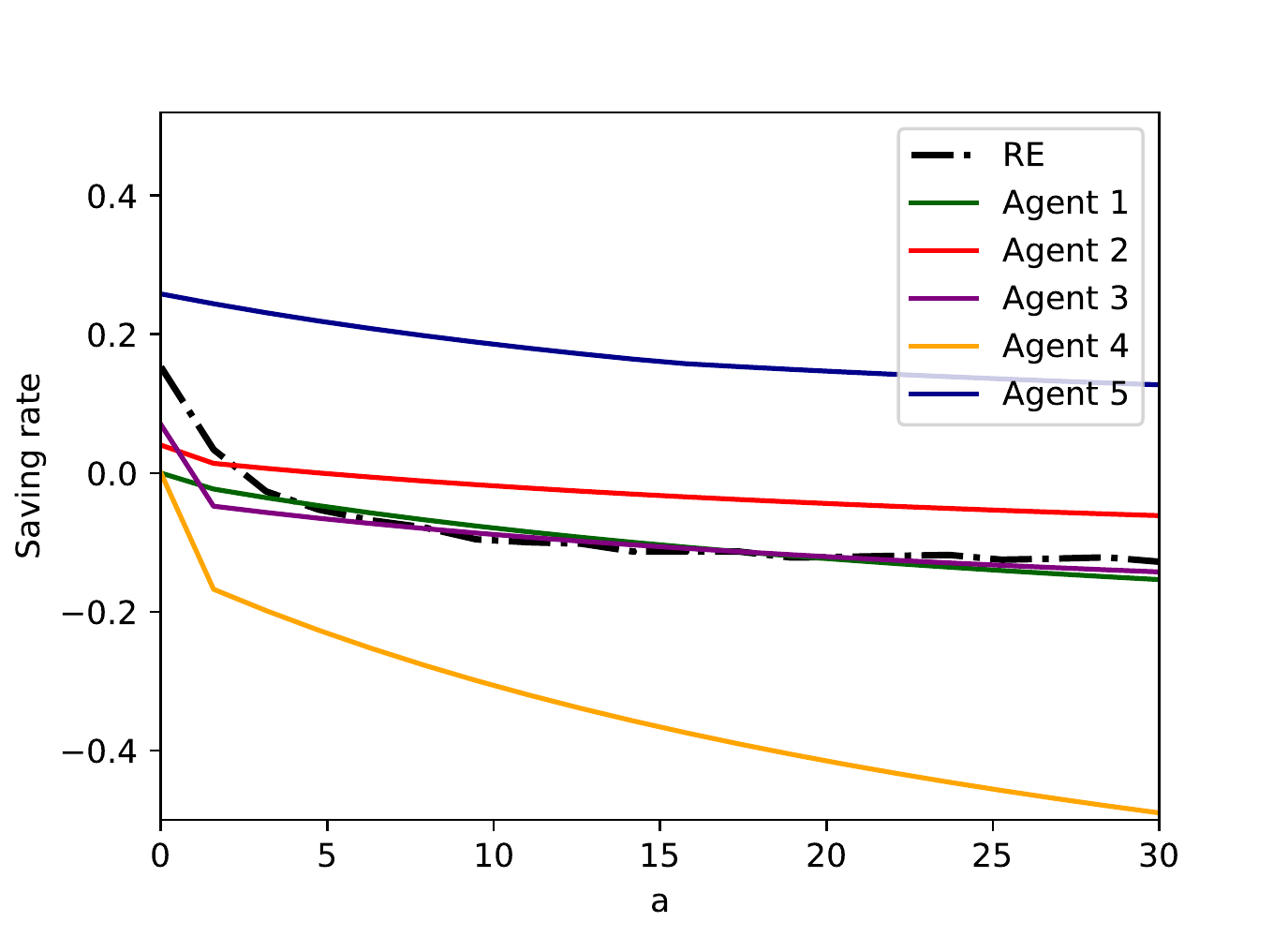} \label{fig:policies}}
    \caption{Illustration of learning outcomes.}\label{fig:patterns}
      \end{center}
\end{figure}

I simulate a large number\footnote{There are between 10,000 and 30,000 NN agents in each simulation. One simulation requires between 5 and 20 hours when performed on a laptop.} of neural network agents over their lifetime. Figure~\ref{fig:patterns} illustrates the model and some of the common learning outcomes. 

For Figure~\ref{fig:timing}, I simulate alternative wealth trajectories for all NN agents in the economy. Those are the trajectories they would have under the same productivity shocks but if they followed the RE policy instead of the policies they learned. I calculate the difference between these trajectories and the actual trajectories. Figure~\ref{fig:timing} shows how the mean and variance of diversion evolve over time. For the first 20 years, decisions are made by rational parents, and there is no diversion. Since childhood is over, agents start to deviate from the RE paths. The dispersion is growing over time, as diversions accumulate. The mean diversion is positive, as NN agents tend to have more savings on average than RE agents.

Figure~\ref{fig:traject} presents examples of realized wealth paths for five NN agents and one RE agent. There are two sources of variation in outcomes. The first one is standard for heterogeneous agents models: the agents had different productivity histories $z_t$ over their lifetimes. In addition to that, for NN agents the difference in outcomes also comes from the difference in learned decision rules. Decision rules are different because agents learned from different experiences and because of the randomness in learning.\footnote{Neural network agents have different random initializations of their networks, different random batch draws at any learning step, and different random external experiences.}

The trajectories of agents 1, 2, and 3 seem to have a nature similar to that of the presented RE trajectory. In contrast, agents 4 and 5 are in a \textit{learning trap}. For the last half of her life, agent 4 has no savings at all. The trap of agent 5 is very different in terms of her consumption-saving behavior. This agent steadily increases the amount of her savings well above levels that rational agents are likely to reach.

The process of learning can be illustrated by the average Euler equation error the same agents got for the experiences in $E$ when they were updating their policies. As can be seen in Figure~\ref{fig:loss}, all agents have relatively large errors at the first steps of their lives. However, agents 1, 2, and 3 successfully update their policies, and over time their Euler errors start to fluctuate around 0. Agent 4, however, is in a save-nothing learning trap. Her policy tells her to save below the liquidity constraint, her policy gradient is zero, she is not able to update her policy, and she is constantly making consumption errors. Agent 5, despite being in the ``trap" of saving too much, still has small Euler errors as her consumption is large. Because her Euler error is small, her learning updates are small, which prevents her from quickly fixing her consumption behavior.  

Finally, the difference between the agents can also be understood by looking at their policy functions.\footnote{To illustrate policies, I plot the saving rate of agents. This is for practical reasons, as it is hard to see any difference if plain policy values are to be plotted (see Figure~\ref{fig:asymp_rat} for an example).} 
Agents 1, 2, and 3 learned policies that are close to the optimal one. In particular, they correctly learned to save less when their wealth is larger. These agents might save from 5 to 15 percentage points less or more than is optimal, which intuitively seems to be a plausible scale of irrationality.
In contrast, agent 4 felt into a learning trap. She has a policy that implies negative saving rates for all values of $a$. Her neural network has output (at least for wealth 0, where she is) that is below the liquidity constraint. She can neither implement this choice nor learn from it, as she always has a zero policy gradient. She is permanently stuck with no savings. Finally, agent 5 learned an insensible policy of constantly increasing her savings.  

\subsection{Policies and wealth}
 
\begin{figure}[t]
    \begin{center}
    \subfigure[Agents with small wealth.]{\includegraphics[width=0.31\textwidth]{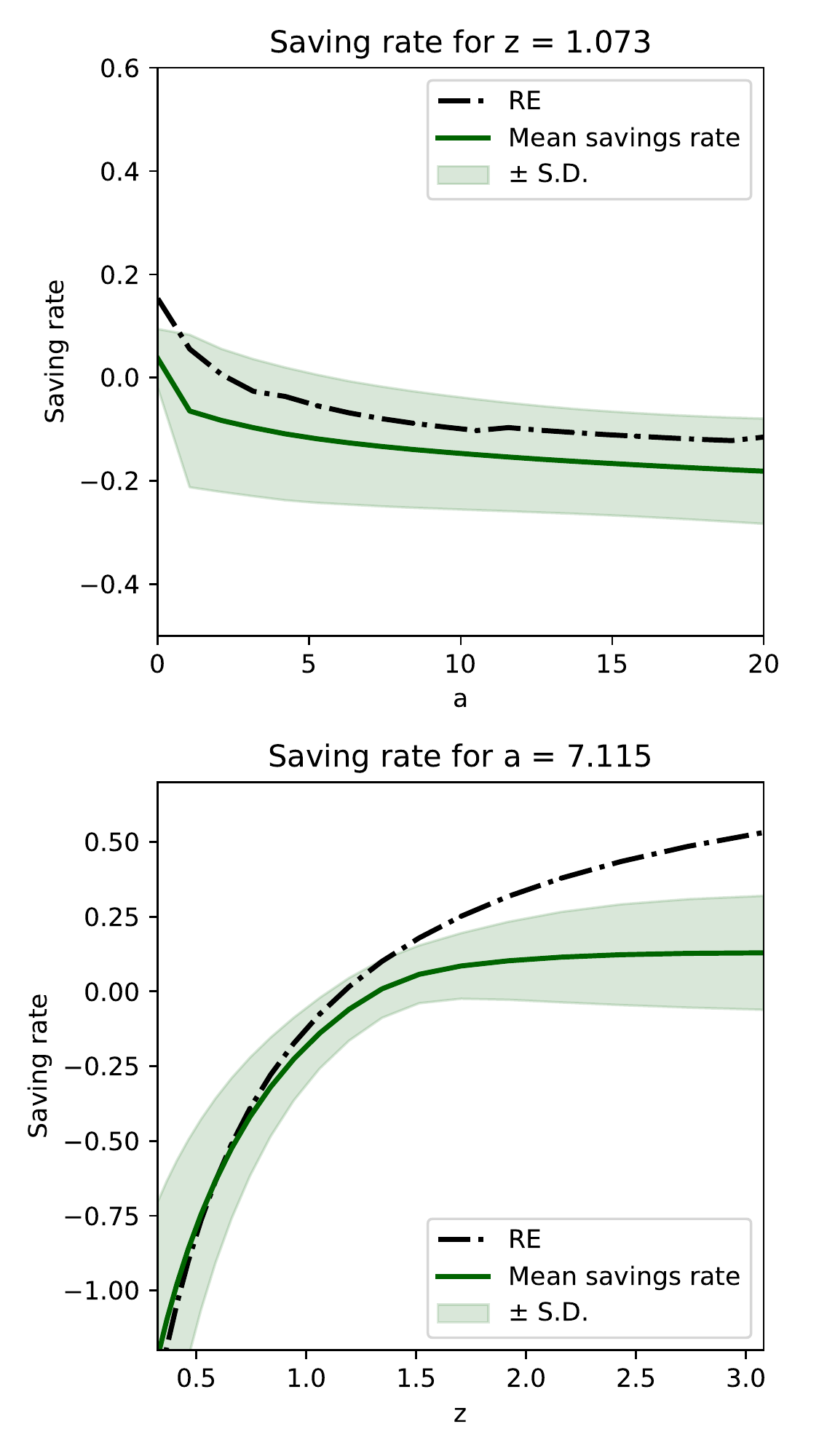} \label{fig:avg_poor}} 
    \subfigure[Agents with medium wealth.]{\includegraphics[width=0.31\textwidth]{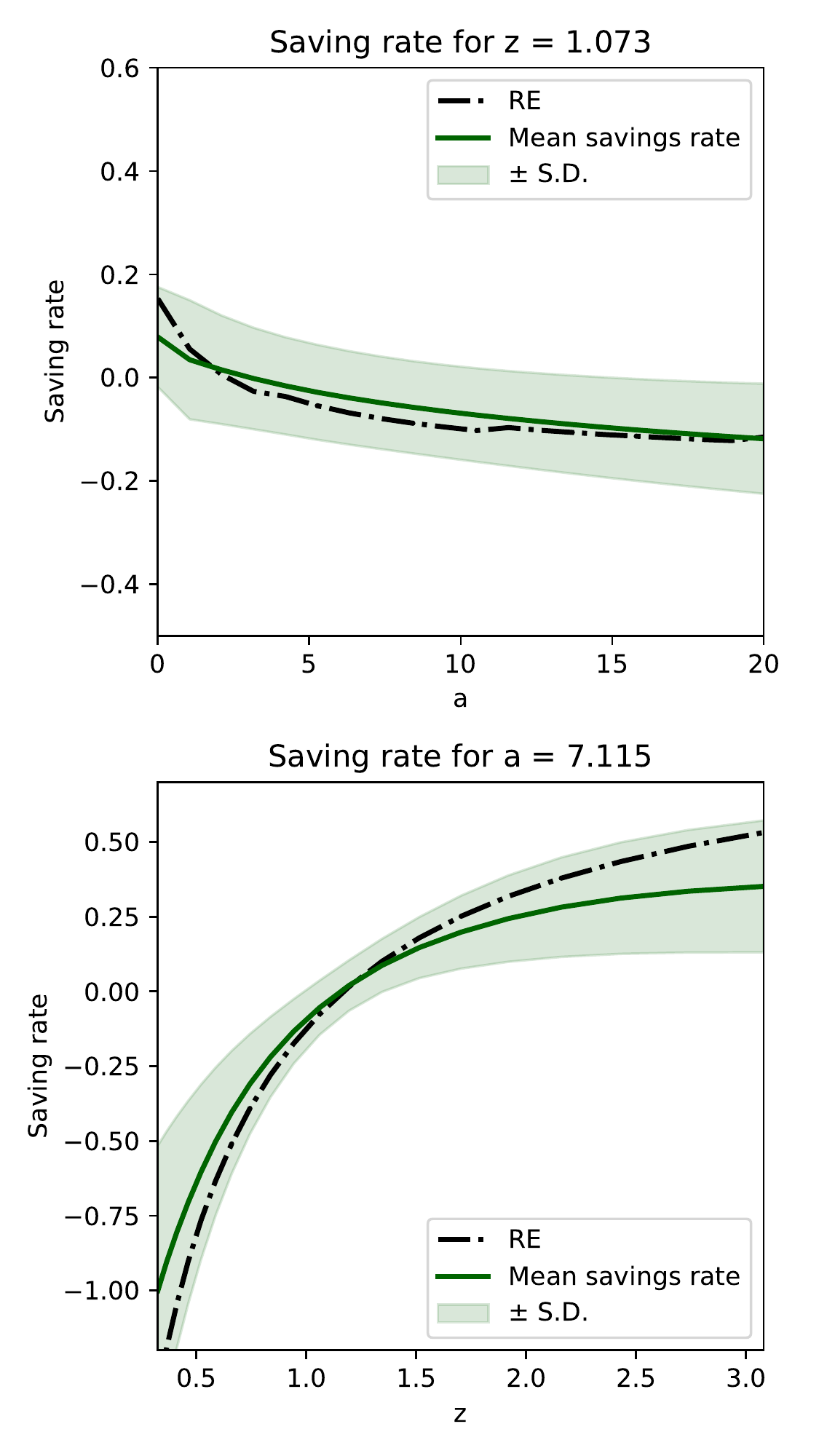} \label{fig:avg_middle}}
    \subfigure[Agents with large wealth.]{\includegraphics[width=0.31\textwidth]{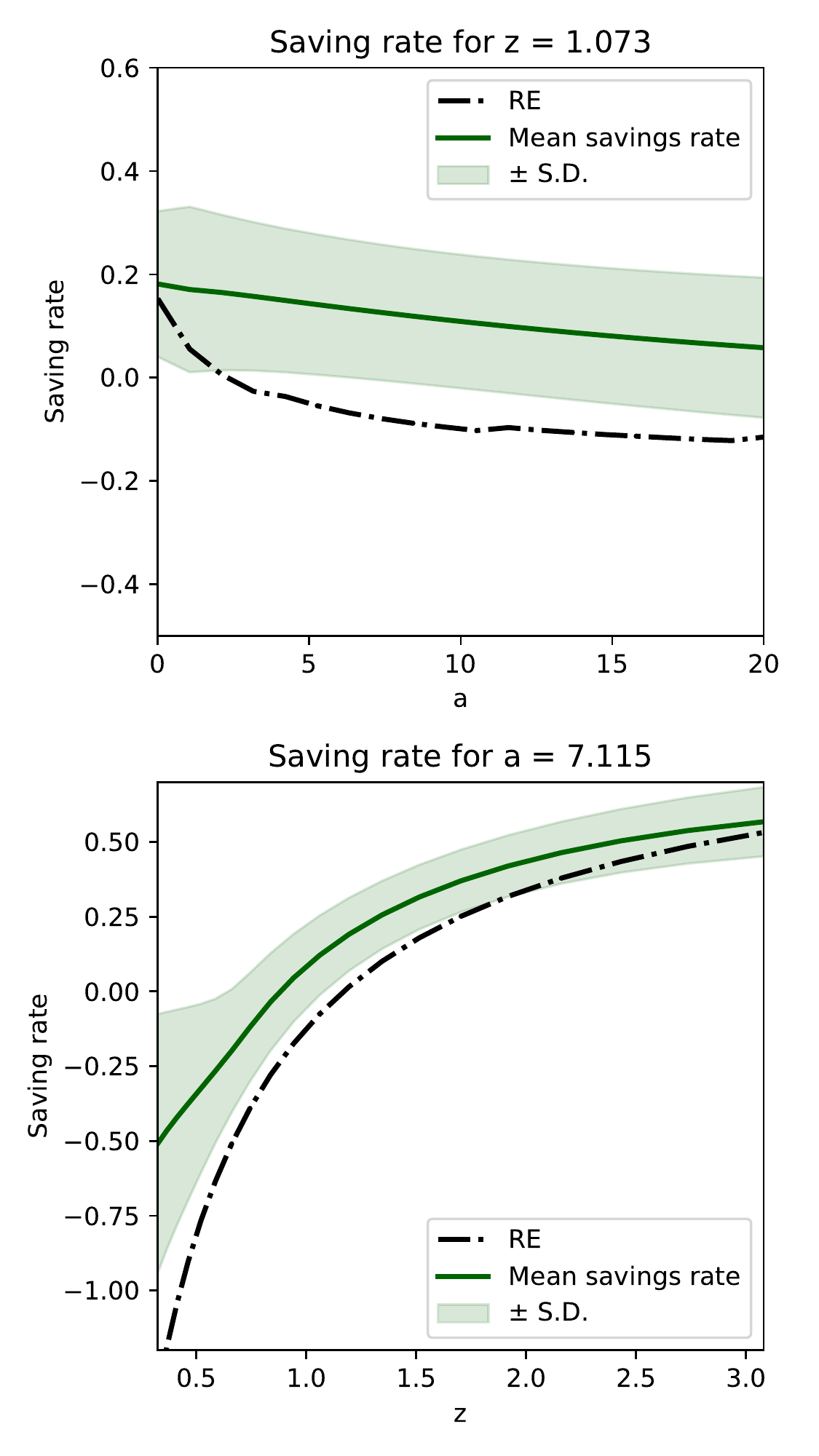} \label{fig:avg_rich}}
    \caption{Average policies by wealth.}\label{fig:avg_pol_3} \medskip
\small Saving rate is defined as $\frac{a_{t+1}-a_t}{ra_t + wz_t}$. When $a$ is varied, $z$ is fixed at the average level. When $z$ is varied, $a$ is fixed at the average level. Width of shaded area equals $2 \times$(standard deviation of policies).
      \end{center}
\end{figure}

To provide a more general illustration of how policies drive wealth in the economy, I split agents into three groups. One group includes all agents with last-period wealth below the 5th percentile of wealth in the RE economy. Another group includes all agents with last-period wealth above the 95th percentile of wealth in the RE economy. The last group includes everyone else. Figure~\ref{fig:avg_pol_3} presents the average of the policies for each of the groups.

As Figure~\ref{fig:avg_poor} shows, agents with low wealth have policies that are biased in the direction of under-saving. Many of them, just like agent 4 from Section~\ref{subsec:examples}, have a policy that implies a negative saving rate everywhere, and are stuck with zero savings. 
In contrast, agents from the medium-wealth group have policies that are approximately unbiased on average, as can be seen in Figure~\ref{fig:avg_middle}. Finally, agents with high wealth significantly over-save, as can be seen in Figure~\ref{fig:avg_rich}. As will be discussed in Section~\ref{subsec:role_of_exp}, their biased policies drive their high wealth.

\subsection{Familiarity with the state space} \label{subsec:role_of_exp}

\begin{figure}[t]
    \begin{center}
    \subfigure[Value of $z$ fixed at the mean level.]{\includegraphics[width=0.48\textwidth]{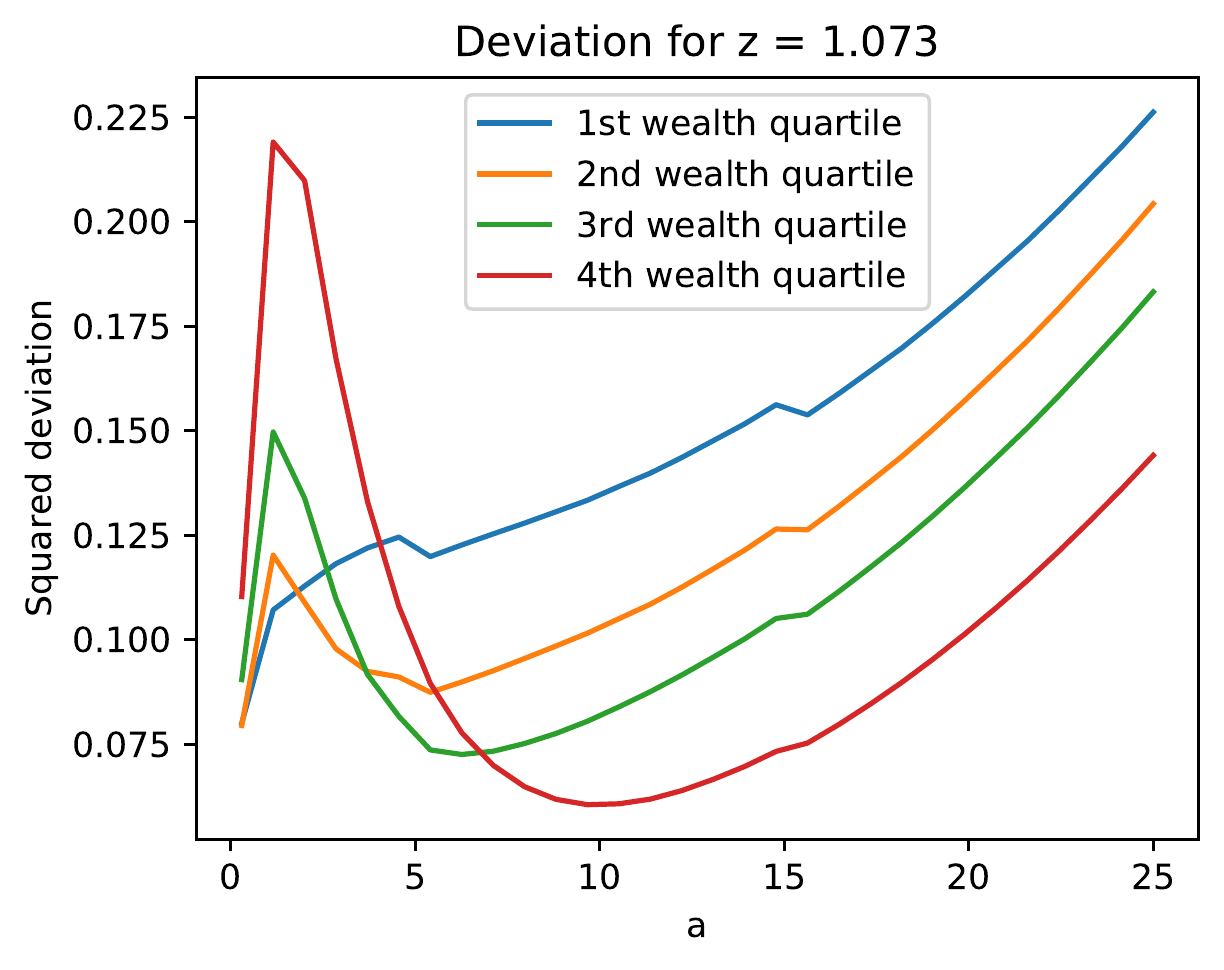} \label{fig:sq_dev_a}} 
    \subfigure[Value of $a$ fixed at the mean level.]{\includegraphics[width=0.48\textwidth]{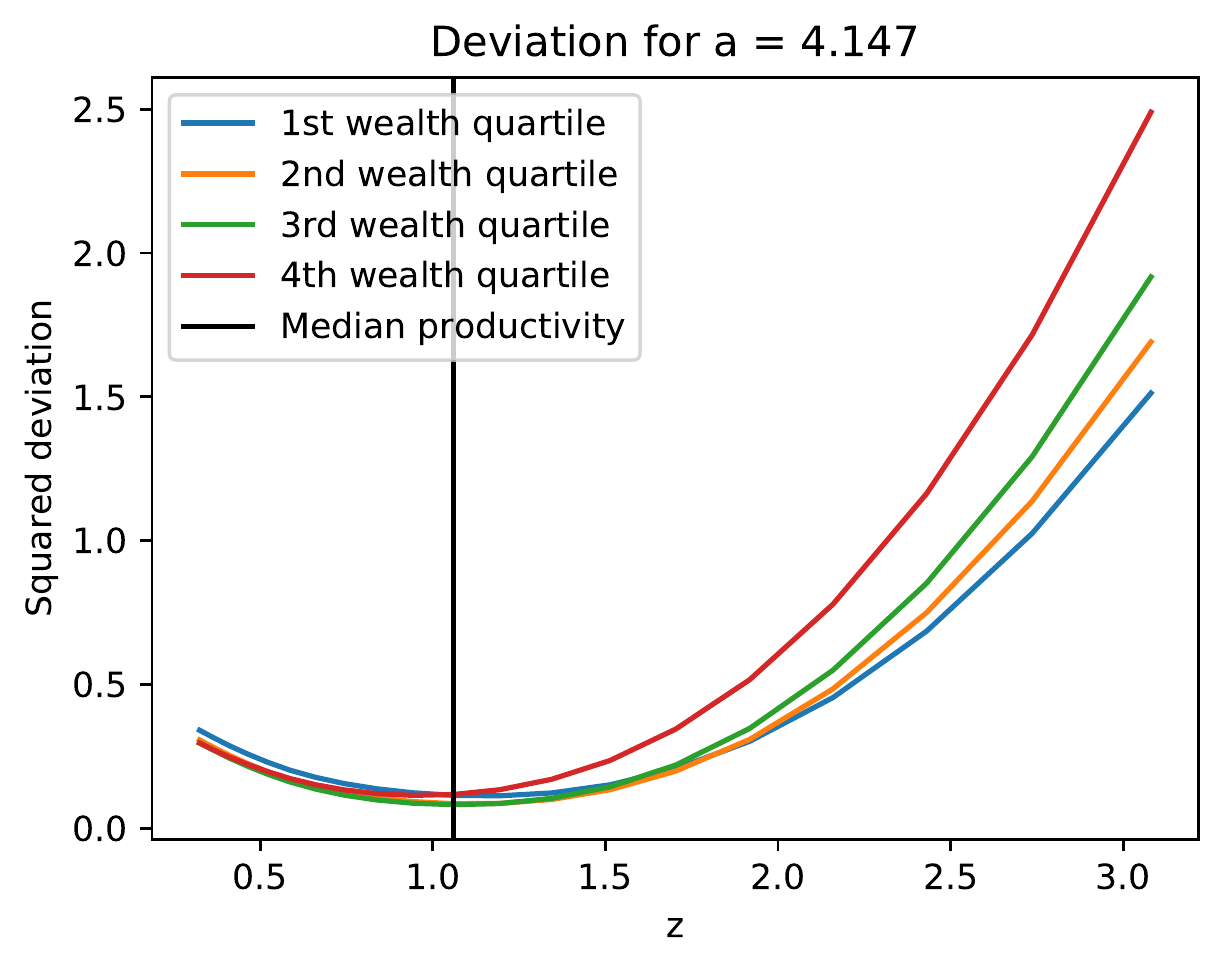} \label{fig:sq_dev_z}}
    \caption{Mean squared difference from RE policy. }\label{fig:sq_dev} 
     
      \end{center}
\end{figure}

The previous exercise shows that there is a statistical association between being rich and saving too much or having low wealth and saving too little. Because there is feedback between policies and experiences, the direction of causality can be unclear. For example, does wealth make the wealthy make some particular over-saving mistakes, or did they become wealthy because of the tendency to over-save? To separate the effect of experiences, I consider the policies that agents have just after their childhood is over. Because before that agents only learned from the experiences generated by their rational parents, all the difference in wealth is independent of their policies and is essentially exogenous. In Figure~\ref{fig:sq_dev}, I plot the average squared difference\footnote{Squared difference from the RE policy is equal, up to a multiplier, to the second-order approximation to the utility loss from choosing a suboptimal action at a particular state.} from the RE policy as a function of the state. 

I split agents into four groups by their wealth quartiles. As can be seen from Figure~\ref{fig:sq_dev_a}, for high values of $a$ the average difference from rationality is larger for agents with smaller wealth. Those agents learned how to act in the region of low savings, and their policies are such that they would make a lot of mistakes if they suddenly became rich. However, the same happens with the wealthiest agents in the region of low wealth, as they would make bad choices there. Because for small values of $a_t$ the range of possible choices of $a_{t+1}$ is small, mechanically all squared differences are small too. However, for some agents these mistakes might have important consequences. As will be discussed in Section~\ref{subsec:mobility}, I find that descendants of 
 parents with low wealth are more likely to accumulate wealth excessively. Descendants of 
 parents with high wealth, on the contrary, are more likely to have no savings in the last period. 

Similarly, Figure~\ref{fig:sq_dev_z} shows a noticeable U-shaped pattern of mistakes that is centered roughly around the median value of productivity $z$. Agents make larger mistakes at the states of productivity that are rarely realized. That happens because those states are largely unknown to the agents. 
I conclude that for neural network agents having experience from a particular region of the state space helps to make better choices there.

\subsection{Reaction to productivity shocks}\label{subsec:z_shocks}
Another class of learning failures I find is inattention to productivity shocks. Some agents fail to learn to react either to all changes in productivity or to productivity shocks with high and rare values. In the model, this is equivalent to passively consuming all labor income after some income threshold. This generates excess sensitivity of consumption in the economy, which will be discussed in Section~\ref{subsec:c_sens}.

\begin{figure}[t]
    \centering
    \subfigure[Examples.]{\includegraphics[width=0.48\textwidth]{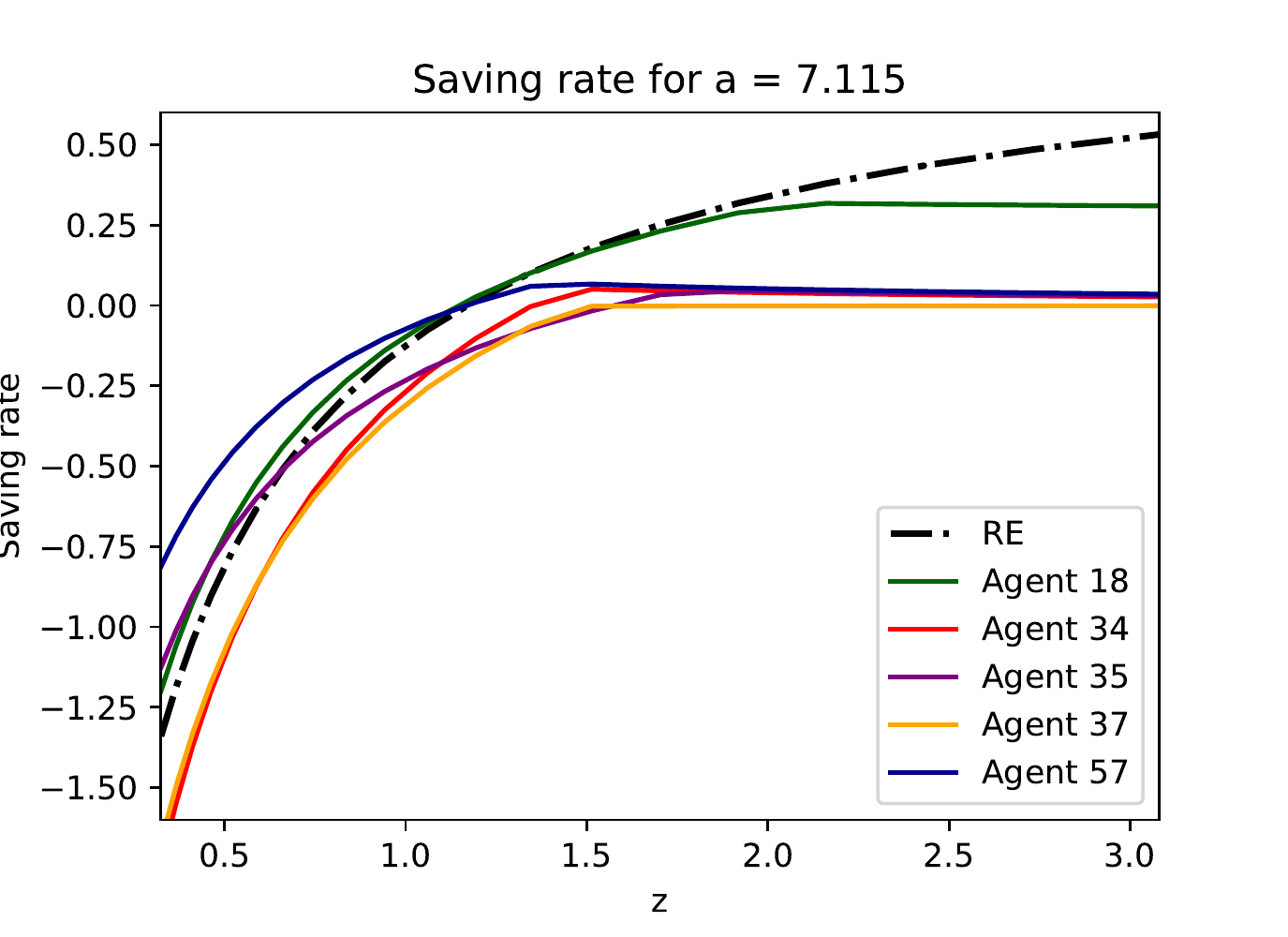}\label{fig:no_z_example}} 
    \subfigure[Distributions of policies.]{\includegraphics[width=0.48\textwidth]{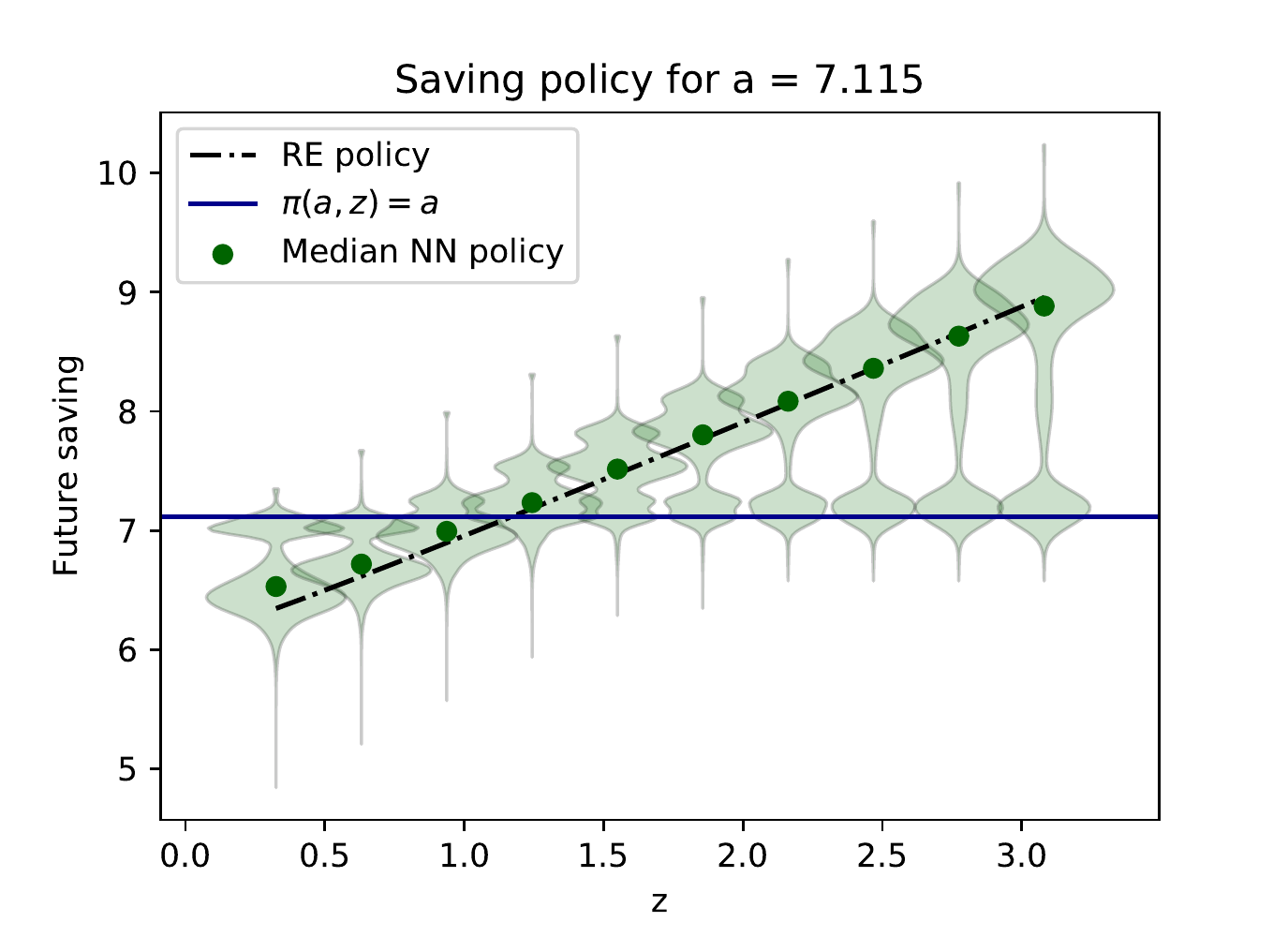}\label{fig:z_pol_violin}} 
    \caption{Failing to learn reaction to shocks.}
    \label{fig:dist_base}
\end{figure}

Figure~\ref{fig:no_z_example} presents several examples of how policies might look in the case when agents do not react to high shocks only. In Figure~\ref{fig:z_pol_violin}, I plot the distributions of the policies for several values of $z$. As can be seen, median policies are quite close to the optimal decision rule, but there is always a mode around the inertial policy that corresponds to the agents who ignore shocks. Thus, agents who are inattentive to shocks account for a significant part of the under-savings biases that can be seen in the bottom row of Figure~\ref{fig:avg_pol_3}. In particular, the bias is strongest in Figure~\ref{fig:avg_poor}, where agents with low wealth under-save a lot on average for high values of $z$. For middle-wealth agents, inattention occurs less frequently, and it is essentially absent for agents with high wealth (this can also be seen if I reproduce Figure~\ref{fig:z_pol_violin} for different wealth groups).

Intuitively, some agents might fail to start thinking about the volatility of shocks. They are stuck around the inertial initialization or update their policies but only in terms of the dependence on the other state variable $a$. Also, high levels of productivity are too infrequent, and many agents do not have a proper sample to learn in that region of the state space. Though this behavior might depend on the choice of initialization (which assumes that agents completely ignore shocks) and the updating framework, it is also both plausible and quite intuitive that some agents would fail to learn that productivity levels from the Markov process $\Gamma$ are correlated or can be used to smooth consumption. This finding resembles the behavior of agents in \cite{gabaix2016behavioral}. There, agents choose to pay partial attention or no attention at all to some state variables. 

Though it is not modeled here formally, this phenomenon might become especially important in times of high volatility. A considerable fraction of agents in the model react to large shocks incorrectly and in the same way. This can have a significant effect on the economy if there is a structural change that makes large shocks more likely. In addition to that, an inability to learn any reaction to shocks might have macroeconomic consequences in a model with a richer structure, say, with monetary and fiscal shocks. If many agents fail to account for the shocks, this can affect the persistency of GDP cycles and the efficiency of monetary or fiscal policy. 


\subsection{Cost of irrationality}
Making suboptimal saving choices leads to smaller expected lifetime utility. How much does an agent who is born irrational lose ex-ante? I find the compensating variation $CV$ as the solution for
$$\mathbb{E} \Bigg[ \sum_{t=20}^{100} \beta^t u\big(c_{i,t}^{RE}\big) \Bigg] = \mathbb{E} \Bigg[  \sum_{t=20}^{100}  \beta^t u\big((1+CV)c_{i,t}^{NN} \big)\Bigg],$$ where $c_{i,t}^{RE}$ and $c_{i,t}^{NN}$ stand for consumption of RE and NN agents respectively. The compensating variation is the fraction that should be added to the consumption of NN agents to make their expected adulthood utility the same as that of RE agents.\footnote{For simplicity, I ignore the value of consumption after the last period (though agents take it into consideration). It should not significantly affect the results as $\beta^{100}$ is a very small number.} Similarly, I find the equivalent variation $EV$ as the solution for
$$\mathbb{E} \Bigg[ \sum_{t=20}^{100} \beta^t  u\big((1-EV)c_{i,t}^{RE}\big)  \Bigg] = \mathbb{E} \Bigg[  \sum_{t=20}^{100}  \beta^t   u\big(c_{i,t}^{NN} \big)\Bigg].$$ This is the fraction that RE agents would be ready to give up ex-ante to keep being rational. 

 Table~\ref{table:cost} presents the results for the two types of NN agents. Not surprisingly, having low rationality is harmful for utility. The cost of that is more than one-tenth of consumption. Having high rationality leads to a smaller disadvantage relative to full rationality, with the loss roughly equivalent to 4\% of consumption. 

\begin{table}[H]
\begin{center}
\begin{tabular}{cccc}
\hhline{~===}
 &  Low rationality & High rationality \\ \hline
CV                  &          13.6\%          &       4.3\%           \\
EV                     &           12.0\%         &       4.2\%           \\ \hline \hline
\end{tabular}
\end{center}
\caption{Equivalent and compensating variations for NN agents vs. RE agents.}
\label{table:cost}
\end{table}

\section{Macroeconomic Implications} \label{sec:macro}
\begin{figure}[t]
    \centering
    \subfigure[Low rationality.]{\includegraphics[width=0.48\textwidth]{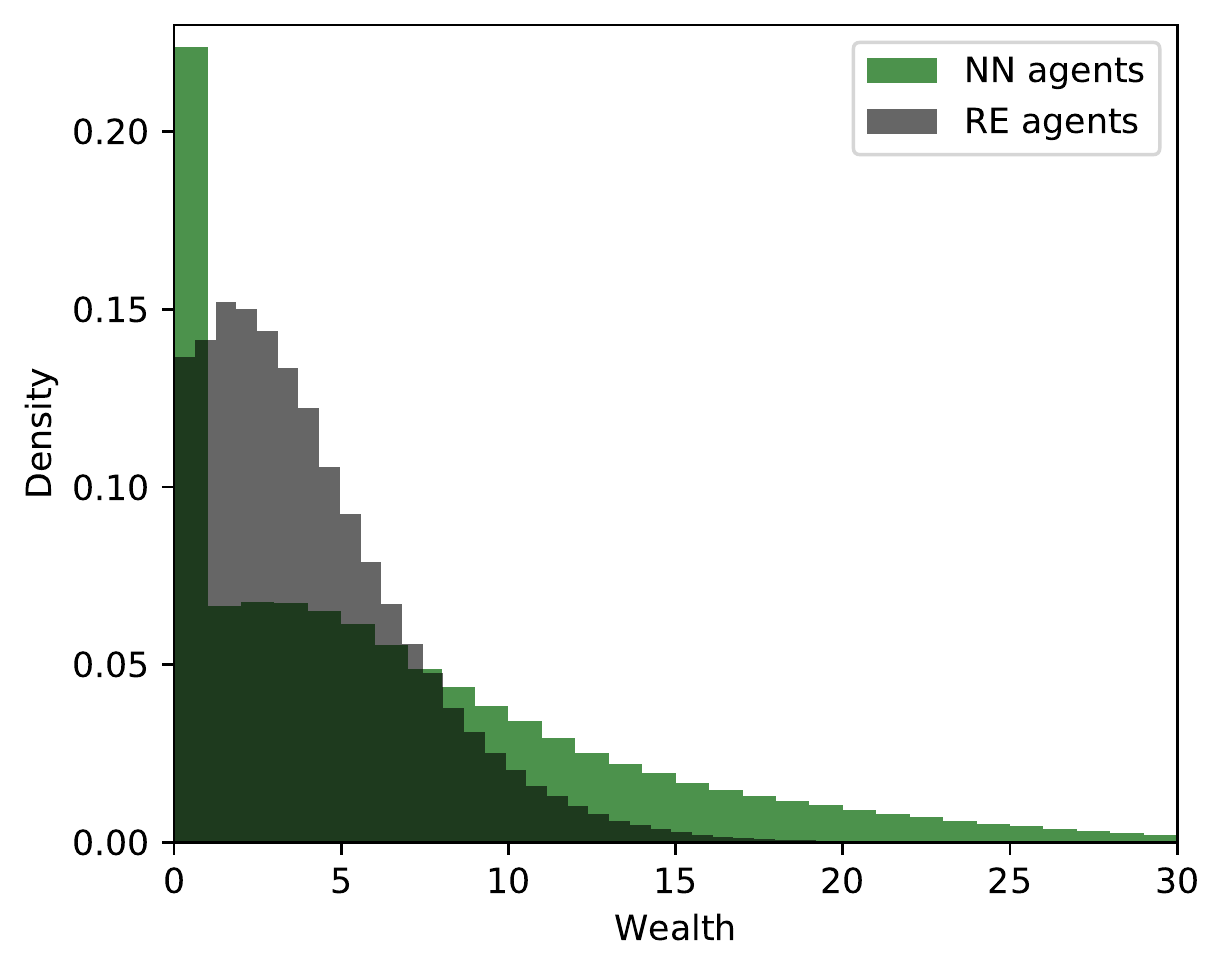}\label{dist_low}} 
    \subfigure[High rationality.]{\includegraphics[width=0.48\textwidth]{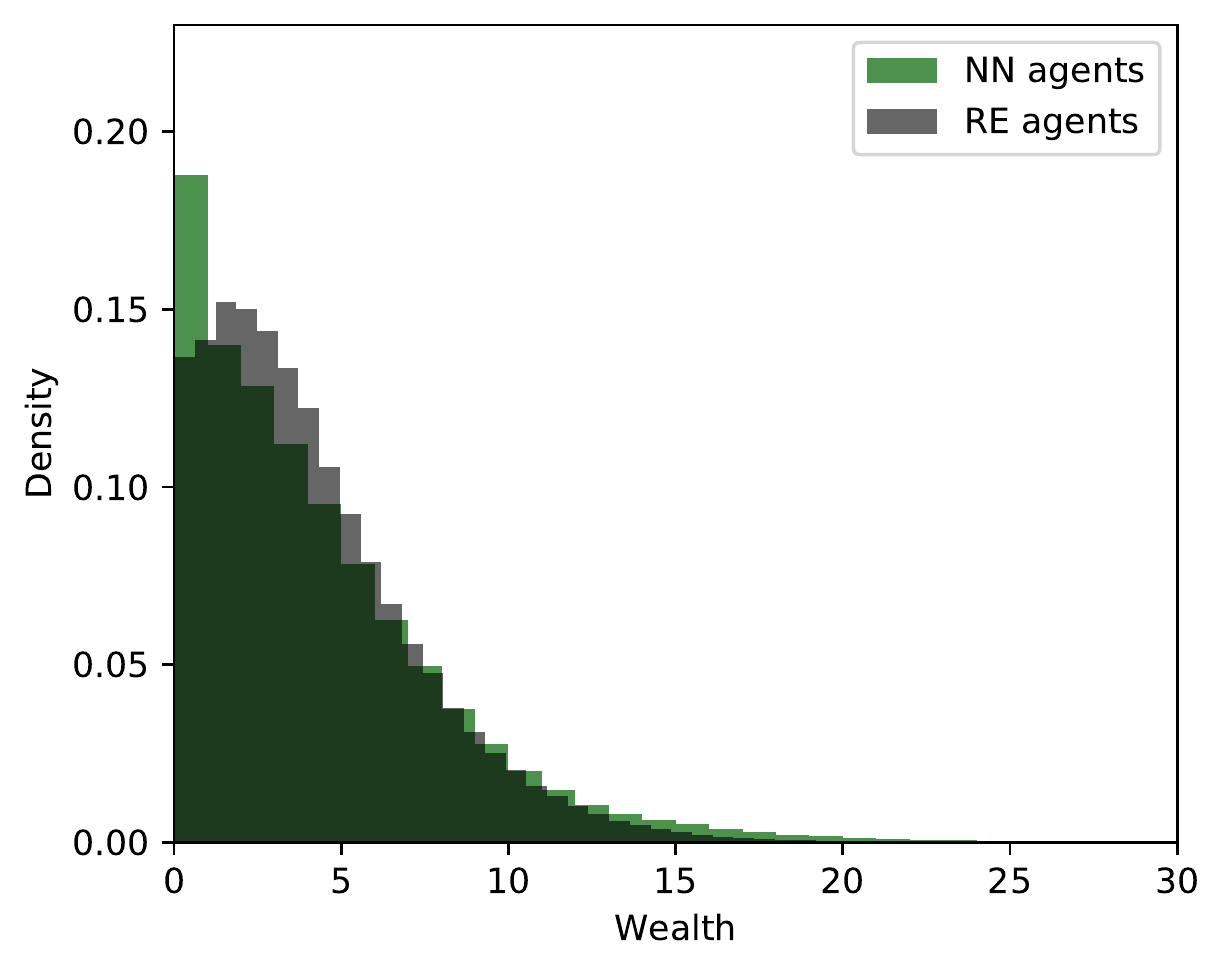}\label{dist_high}} 
    \caption{Distribution of savings in the economy.}
    \label{fig:dist_base}
\end{figure}

 Figure~\ref{fig:dist_base} shows the simulated distributions of savings for rational agents and the two types of neural network agents. As can be seen in Figure~\ref{dist_high}, under high rationality the results closely approximate the outcomes under rational expectations. However, as is clear from Figure~\ref{dist_low}, learning and bounded rationality might have important effects on the macroeconomy.
 
\subsection{Hand-to-mouth agents} Models with liquidity constraints typically do not generate a mass of agents around the minimum wealth. This happens because rational agents save precautionarily in order not to find themselves in the position where they would not be able to borrow to smooth their consumption. However, in the real world many households have essentially no savings. Following \cite{zeldes1989consumption}, I say that an agent has hand-to-mouth status if her wealth is smaller than her labor income over two months.\footnote{Formally, agent $i$ has hand-to-mouth status at period $t$ if $a_{i,t} < \frac{wz_{i,t}}{6}$.}
In the rational expectations equilibrium, I find that only 2.6\% of agents are hand-to-mouth. In contrast, \cite{aguiar2020hand} report that 22.7\% of the households surveyed for the Panel Study of Income Dynamics (PSID) are hand-to-mouth according to this definition. Similar to \cite{ilut2020economic}, I test whether policy learning can explain the discrepancy between the data and the model with rational agents.

As can be seen from Figure~\ref{fig:dist_base}, neural network agents have a tendency to accumulate around the lower bound for assets. The first row of Table~\ref{table:h2m} presents the fraction of hand-to-mouth agents in the simulation.
Both types of NN agents are substantially more likely to become hand-to-mouth than agents who comply with rational expectations. As before, the configuration with low rationality delivers larger differences. In the model, hand-to-mouth NN agents most often are in a save-nothing learning trap. This suggests that behavioral patterns related to  trap-like learning outcomes might contribute to the large share of hand-to-mouth agents observed empirically. 

\begin{table}[h]
\begin{center}
\begin{tabular}{lccccc}
\hhline{~=====}
& PSID  & Low rationality& High rationality & RE \\ \hline
$\mathbb{P} \big($H2M$_{t}\big)$ &  0.227         &         0.157          &       0.072    &   0.026         \\ 
 $\mathbb{P} \big($H2M$_{t+2} |$   H2M$_{t}\big)$ &  0.648              &       0.897             &         0.686  &  0.511        \\ 
 $\mathbb{P} \big($H2M$_{t+4} |$   H2M$_{t}\big)$ &   0.584             &         0.845           &        0.526    &  0.321      \\ \hline \hline
\end{tabular}
\end{center}
\caption{Frequency and persistence of hand-to-mouth status. Estimates for PSID are from  \cite{aguiar2020hand}.}
\label{table:h2m}
\end{table}
I also calculate the probability that agents stay hand-to-mouth after two periods and four periods in the RE and NN economies. The last two rows of Table~\ref{table:h2m} present the results. 
For the same reason that hand-to-mouth status is rare in RE models, it is also not particularly persistent. After hitting the liquidity constraint, agents usually quickly accumulate some amount of assets. In the data, however, the status is more persistent, and this is also true in the NN economy. The hand-to-mouth neural network agents mostly are the ones in a learning trap, which often means that agents learned to save nothing and do not update policies anymore. Hand-to-mouth status is thus extremely persistent (in fact, the model implies larger persistence than observed in the data).

\subsection{Wealth inequality}
It is hard to fit the empirically observed wealth distribution with a typical heterogeneous agents model, in particular, because we observe thick upper tales in the data (see~\citealp{benhabib2018skewed}, for a review). 
As is clear from Figure~\ref{fig:dist_base}, neural network agents end up with a wealth distribution more dispersed than that of rational agents. Table~\ref{table:inequal} shows that NN agents indeed have various measures of inequality closer to those obtained from the PSID and reported in \cite{krueger2016macroeconomics} than the rational expectations equilibrium. 

\begin{table}[h]
\begin{center}
\begin{tabular}{lccccc}
\hhline{~=====}
               & PSID & Low rationality  & High rationality & RE  \\ \hline
Gini           & 0.77                                    &             0.535       &        0.486     &               0.411      \\
Top 1\% share  & 30.9                                          &           5.8         &        5.2    &           3.9       \\
Top 5\% share  & 53.7                                      &             20.1       &           18.2  &             15.2      \\
Top 20\% share & 82.7                                  &            53.5        &            48.6    &           44.2    \\ \hline \hline
\end{tabular}\end{center}
\caption{Wealth inequality. Estimates for PSID are from \cite{krueger2016macroeconomics}. Shares of wealth are in percentage points.}
\label{table:inequal}
\end{table}

Broadly, the main reason for higher inequality in an economy of neural networks is that the heterogeneity in thinking adds up to the heterogeneity from idiosyncratic shocks. 
There are two particular extreme cases of how this happens. The first one is the higher share of agents stuck with no savings. Those are the agents who at some point learned a policy that keeps them with no wealth and that does not allow for improving learning. The second one is the high number of agents who save excessively and end up in the upper tail of the distribution. Thus, the upper tail consists of agents who learned a policy that is biased toward high saving. Those agents might have small updates because of high consumption or upward-biased policy updates, or because they are not able to learn further at all. 

Though it seems unlikely that in the real world all of the wealthiest (or all of the least well-off) are the ones who just failed to learn a good decision rule, this finding is still suggestive. I consistently find agents diverging to save-nothing or save-everything policies for different set-ups and different hyperparameters. The main explanations for this are the self-perpetuating biases and learning traps mentioned above. These mechanisms might also be present in the real world, generating extreme learning outcomes and contributing to inequality in the economy. The model predicts a moderate but non-negligible effect of learning on inequality.

\subsection{Marginal propensity to consume}
The baseline rational expectations model fails to deliver an average marginal propensity to consume close to the empirical estimates. The latter tend to vary from 20\% to 60\% (see a summary of the empirical literature in~\citealp{carroll2017distribution}). In contrast, the RE equilibrium has an average MPC around $7.7\%$. I define the marginal propensity to consume out of transitory income as $$MPC(a,z) = 1 - \frac{1}{1+r}\frac{\partial}{\partial a}\pi(a, z),$$
which for a rational agent is the same as the share consumed out of a marginal increase in cash-on-hand coming from a transitory source. 

\begin{table}[h]
\begin{center}
\begin{tabular}{ccccc}
\hline \hline
Data  &  Low rationality  & High rationality & RE \\ \hline
 $[0.2, 0.6]$             &         0.17           &       0.11      &  0.08       \\ \hline \hline
\end{tabular}
\end{center}
\caption{Marginal propensity to consume. The range of empirical estimates is from \cite{carroll2017distribution}.}
\label{table:mpc}
\end{table}

\begin{figure}[t]
    \begin{center}
    \subfigure[Marginal propensity to consume in simulation, \newline by wealth.]{\includegraphics[width=0.48\textwidth]{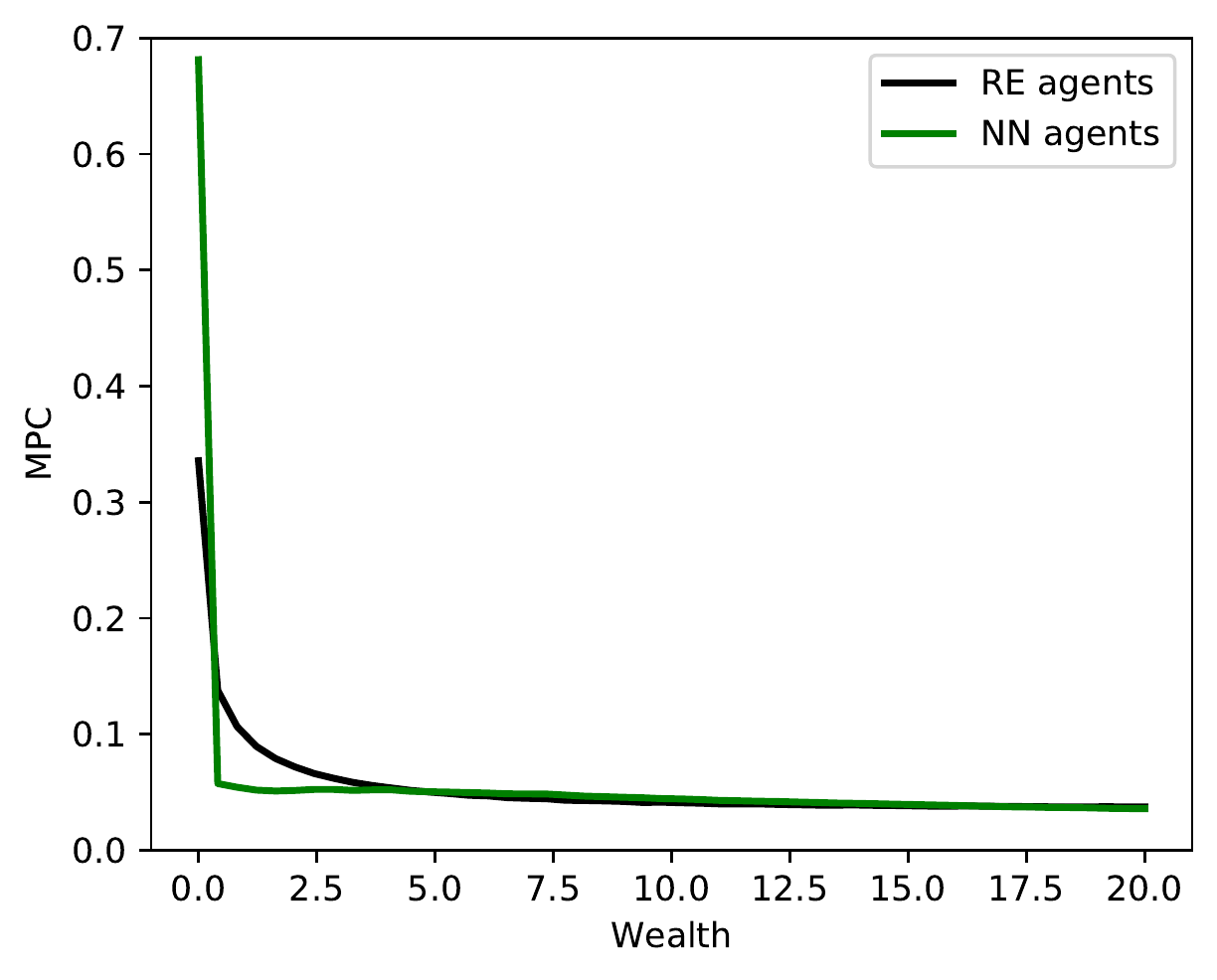} \label{fig:mpc}} 
    \subfigure[Elasticity of changes in consumption $\Delta \log c_{i,t+1}$ with respect to predictable changes in income $\mathbb{E}_t \Delta \log y_{i,t+1}$, by wealth.]{\includegraphics[width=0.48\textwidth]{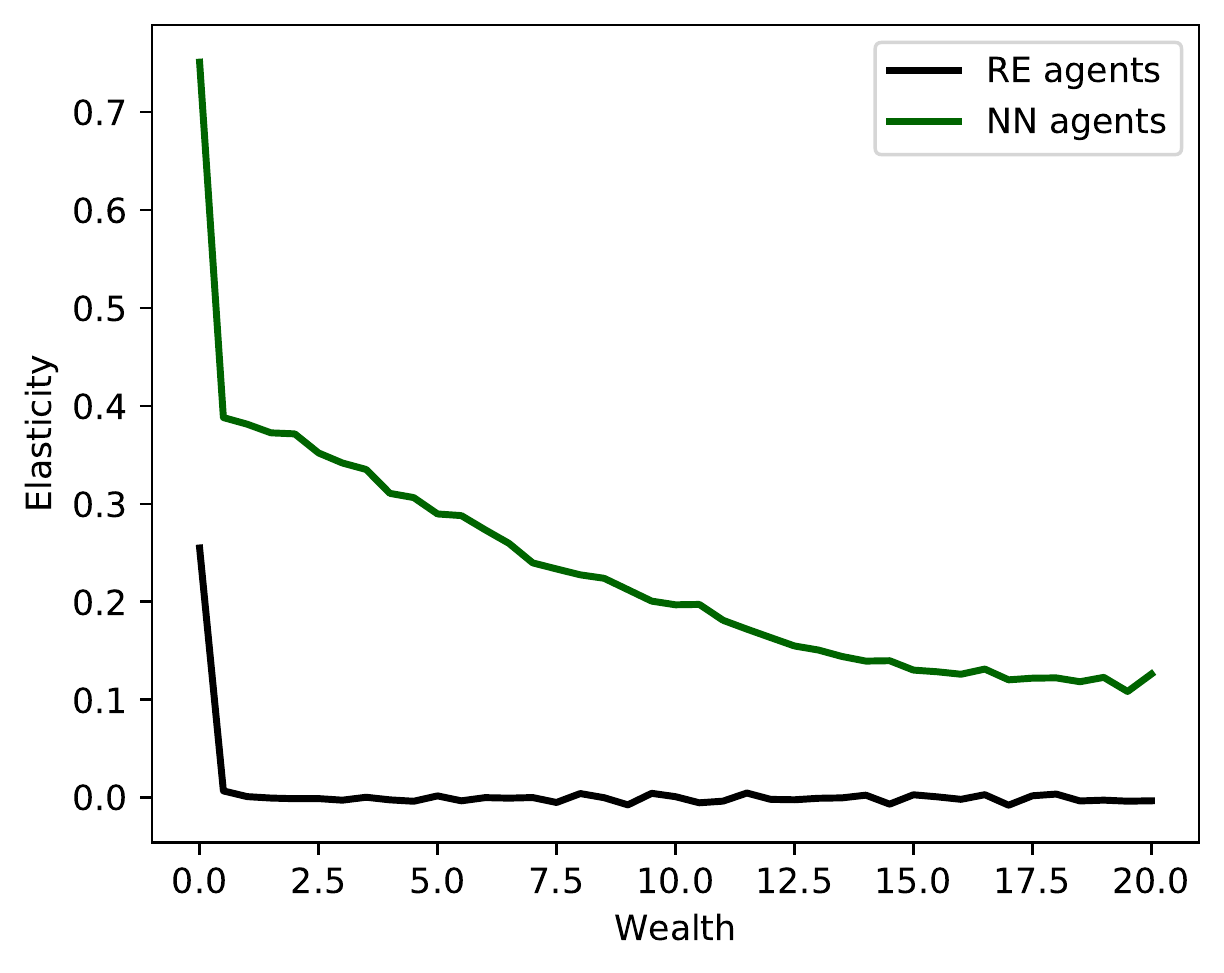} \label{fig:elasticity_cons}}
    \caption{Consumption behavior of NN agents.}
      \end{center}
\end{figure}

I calculate the average marginal propensity to consume in the simulated data. Table~\ref{table:mpc} shows that the economy with low rationality has an average MPC roughly twice that of the Aiyagari economy with rational agents.
Further, I split the generated data into brackets by current asset holdings. Figure~\ref{fig:mpc} presents the dependence of the average MPC on wealth. As can be seen, the difference mainly comes from the agents near the liquidity constraint. The NN economy generates a larger mass of hand-to-mouth agents. Those agents have a higher MPC even in the rational expectations model (\citealp{carroll1996concavity}), but in the NN economy those often are the agents in a save-nothing learning trap. The MPC of those agents is extremely high (and usually is equal to 1).

\subsection{Sensitivity of consumption}\label{subsec:c_sens}
Another puzzle against which NN agents can be tested is the ability of expected changes in income to predict future changes in consumption. With perfect capital markets and under the permanent income hypothesis, predictable changes should not matter. One way to test this is to consider the regression $$\Delta \log c_{i,t+1} = \alpha + \beta \log y_{i,t} + \epsilon_{i,t.}$$ If $\log y_{i,t}$ follows an AR(1) process with autocorrelation $\rho$, then $\mathbb{E}_t \Delta \log y_{i,t+1} = -(1-\rho)\log y_{i,t}$, and $-\frac{\beta}{1-\rho}$ is the elasticity of 
changes in consumption with respect to predictable changes in income. This approach was first implemented by \cite{flavin1981adjustment}, who found evidence against the PIH in the data. After that, some studies also found a strong rejection (see~\citealp{zeldes1989consumption, lusardi1996permanent}), while others found limited or no evidence of that (see, for example,~\citealp{blundell2008consumption, hsieh2003consumers}). 

I calculate the elasticity in the simulated data. Table~\ref{table:elast} shows that the NN economy has a much larger elasticity than the Aiyagari economy with rational agents. Further, I split the generated data into brackets by current asset holdings. Figure~\ref{fig:elasticity_cons} presents the dependence of elasticity on wealth. As can be expected, for rational agents, deviations from the PIH happen only around the liquidity constraint. In contrast, NN agents exhibit excess sensitivity of consumption for all values of wealth.

\begin{table}[h]
\begin{center}
\begin{tabular}{ccccc}
\hline \hline
Data  & Low rationality  & High rationality & RE \\ \hline
  [0,0.4]         &         0.339          &         0.063     &  0.026        \\ \hline \hline
\end{tabular}
\end{center}
\caption{Elasticity of changes in consumption $\Delta \log c_{i,t+1}$ with respect to predictable changes in income $\mathbb{E}_t \Delta \log y_{i,t+1}$. Upper bound for empirical estimates is from \cite{lusardi1996permanent}.}
\label{table:elast}
\end{table}

Divergence from rational expectations is likely to make agents fail to perfectly extract the predictable component from labor income fluctuations. One particular mechanism behind the high sensitivity in the economy was described in Section~\ref{subsec:z_shocks}, where I show that some agents do not learn to react to productivity shocks at all, and some ignore increases in productivity after certain thresholds. Those agents can change consumption one-to-one after a change in income. As was discussed, they mostly have low wealth. This is consistent with Figure~\ref{fig:elasticity_cons}, which shows that richer NN agents have a smaller elasticity of consumption.

\subsection{Intergenerational mobility} \label{subsec:mobility}
In this part, I investigate the effect of learning and bounded rationality on intergenerational mobility. There are many channels through which parental outcomes are transmitted to children's outcomes: transfers, bequests, education, ability, health, social capital, norms, behavioral patterns, and so on (see~\citealp{black2010recent}, for a review). Among those channels, two are related to the framework considered in this project. \cite{waldkirch2004intergenerational} found a correlation between the consumption patterns of parents and children in the PSID. \cite{charles2003correlation} use the PSID to investigate the relationship between the child's wealth and parental wealth. They found that a part of the relationship can be explained by parents and children having similar savings propensities, preferences for risk, and investment behavior. All of that suggests that learning related to the consumption-saving problem might play some role in the intergenerational wealth dynamics.

 One reason why mobility might be different for neural network agents is bounded rationality. It affects the lifetime wealth trajectories and might contribute to convergence or further divergence of agents who started from unequal wealth levels. The direction of this effect is ambiguous. For example, if agents just make random consumption mistakes, it would increase mobility as it would introduce noise that makes starting conditions less important. Alternatively, if all agents follow policies close to the inertial rule $\pi(a, z) = a$, then they would preserve their inherited wealth and there would be no wealth mobility at all. 
 Another reason for different mobility comes from learning. In the first periods of life, learning depends not on the experiences generated by an agent but on the experiences generated by her parent. Thus, learning provides an additional channel through which parental outcomes determine the outcomes of the child.

  \begin{figure}[t]
    \begin{center}
    \subfigure[Probability of having extreme wealth.]{\includegraphics[width=0.485\textwidth]{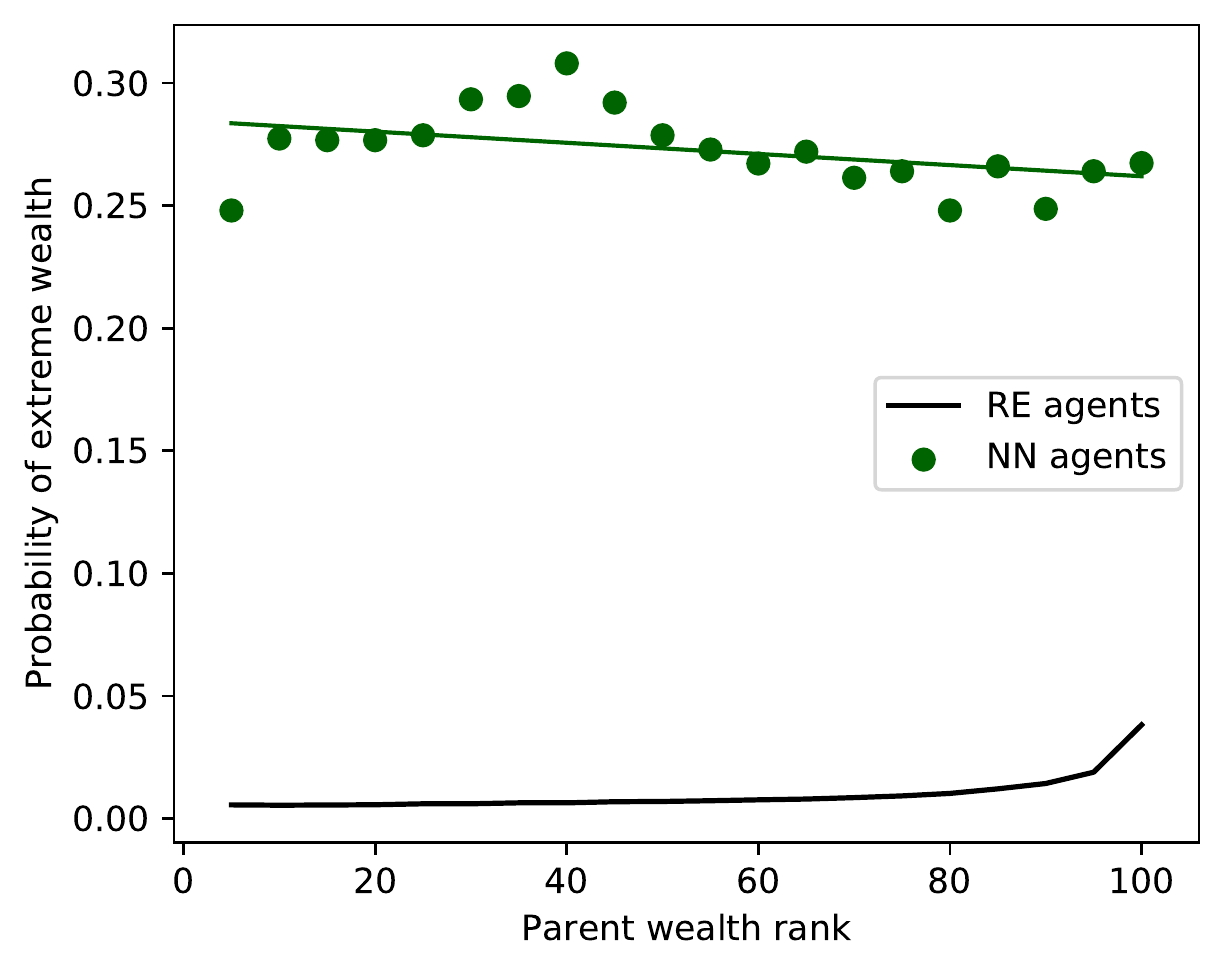}\label{fig:chetty_sr}}    
    \subfigure[Probability of being hand-to-mouth.]{\includegraphics[width=0.485\textwidth]{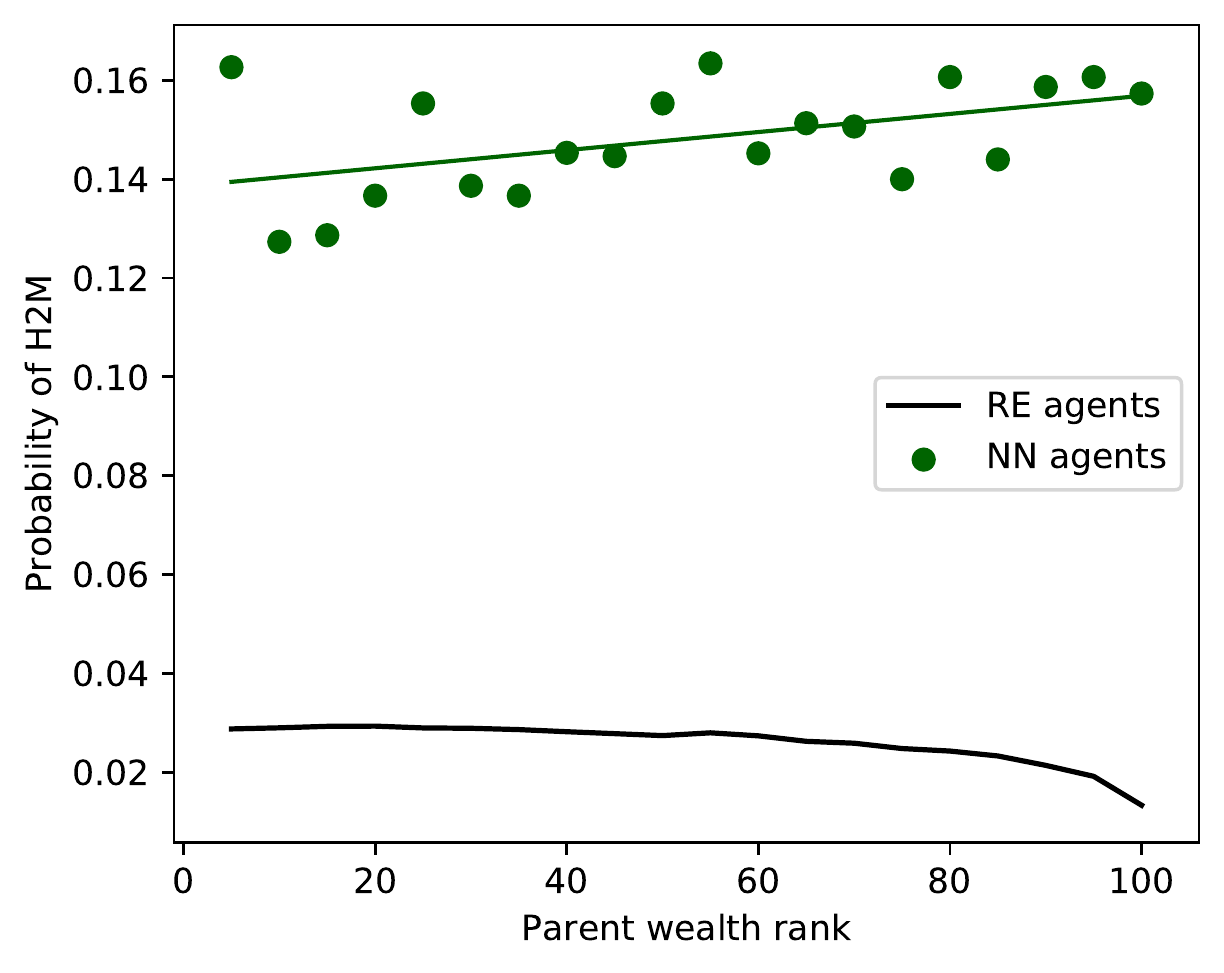}\label{fig:chetty_h2m}}
    \caption{Probability of extreme outcomes vs. parental wealth rank.}\medskip
\small 
Extreme wealth is defined as being above the 99th wealth percentile in the RE economy in the last period. Parents are RE agents. 
\end{center}
\end{figure}

\begin{figure}[t]
    \begin{center}
    \subfigure[RE parents.]{\includegraphics[width=0.485\textwidth]{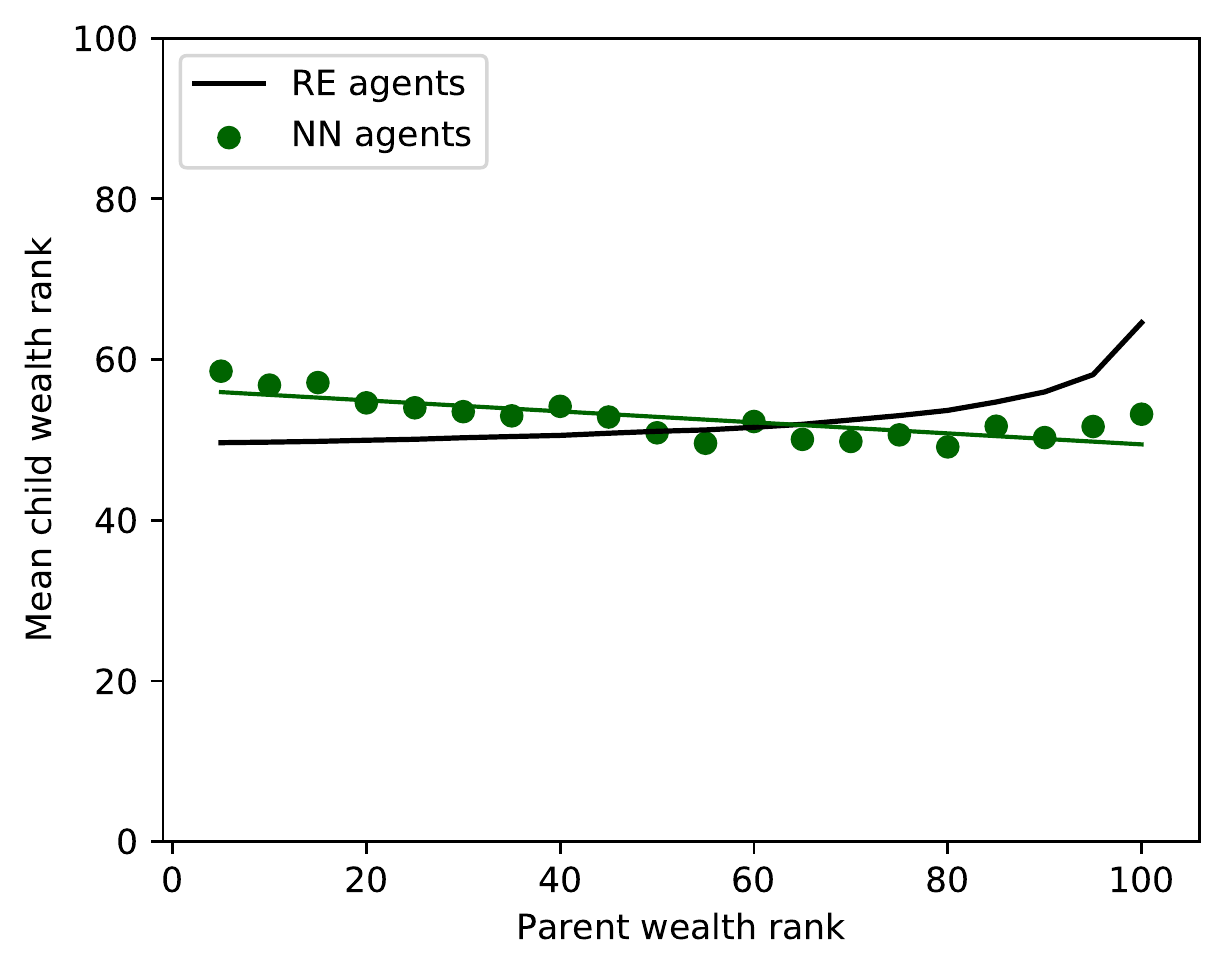}\label{fig:chetty}}
    \subfigure[NN parents.]{\includegraphics[width=0.485\textwidth]{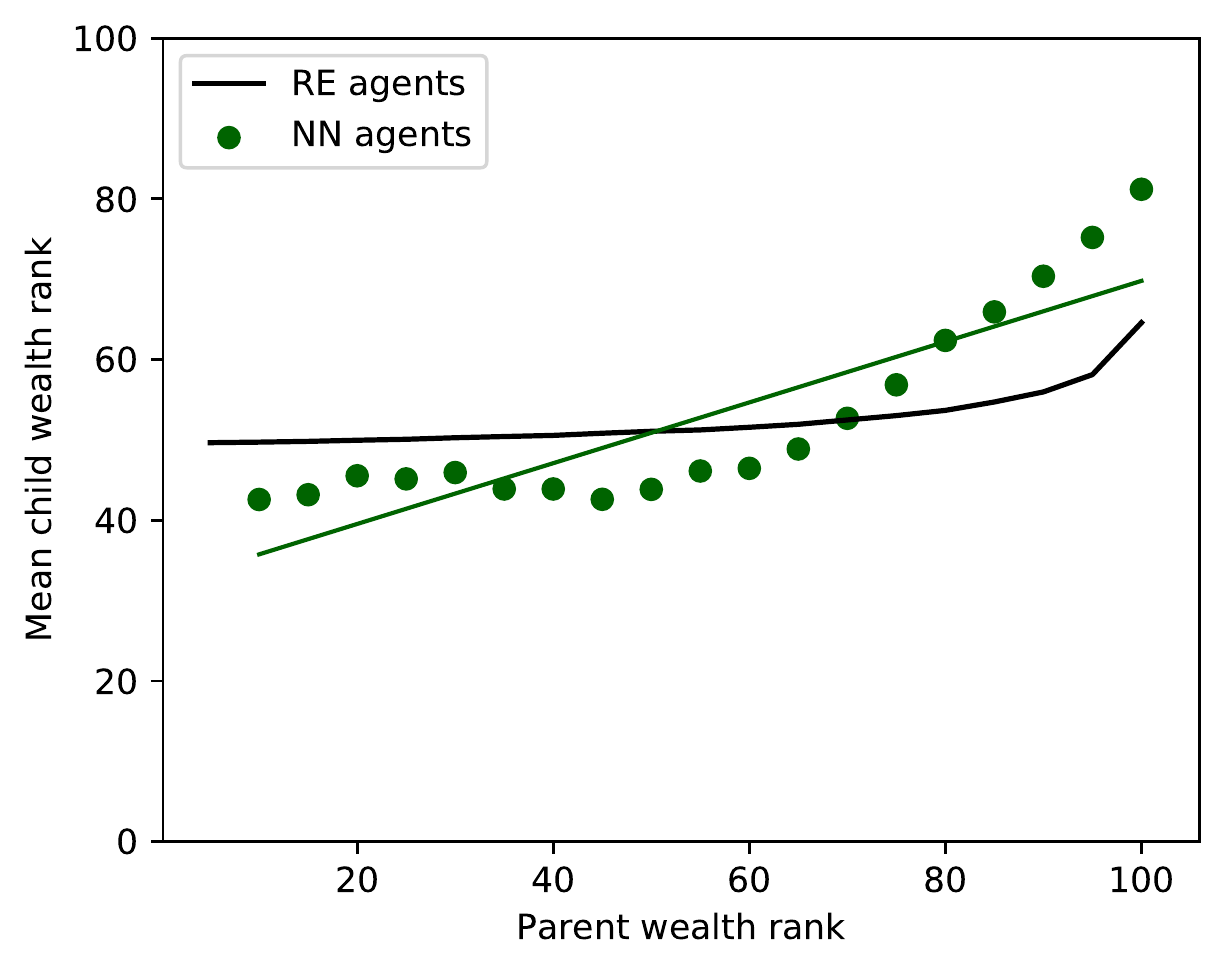}\label{fig:chetty_2nd_gen}}
    \caption{Association between children’s and parents’ percentile wealth rank. The outcomes are in the last period.}
      \end{center}
\end{figure}

I find that learning can either contribute to higher mobility or suppress it. First, let us see the case when learning serves as an intergenerational equalizer. A particularly interesting mechanism for that is the following. In the model, descendants of parents with smaller savings more often accumulate extreme levels of wealth (defined as being above the 99th wealth percentile in the RE economy); see Figure~\ref{fig:chetty_sr}. Similarly, descendants of parents with higher savings more often are hand-to-mouth in the last period, as can be seen in Figure~\ref{fig:chetty_h2m}.
All of that, of course, is not the case for rational agents.\footnote{Here and below, for RE agents I consider the wealth of the same agent but the corresponding number of periods later.} This is likely to happen because of the learning pattern observed in Section~\ref{subsec:role_of_exp}. Those with low-wealth (high-wealth) parents are less aware of the right behavior in the region of high (low) savings, and are more likely to make mistakes if a sequence of shocks brings them there. In addition to this mechanism, consumption mistakes coming from imperfect learning generally might increase mobility as they can work as a noise that distorts the role of initial wealth. All in all, parental wealth has a slightly negative effect on the wealth of neural network agents at the end of their lives. This can be seen in Figure~\ref{fig:chetty}, where I present rank-rank plots in the fashion of \cite{chetty2014united} for both NN and RE agents.

However, this discussion misses another potential channel of how learning might affect mobility. If a parent was not rational as in the model but instead an agent stuck in an under- or over-saving learning trap, then this trap is likely to substantially affect learning by the descendant, and, consequently, her economic outcomes. To see whether this is the case, I simulate the second generation of neural network agents. Those agents learn from the last 20 experiences of their neural network parents before becoming adults, and then inherit their wealth. Figure~\ref{fig:chetty_2nd_gen} shows that this indeed makes wealth much more persistent across generations. In particular, it can be seen that wealthier agents account for most of the effect. A descendant of a wealthy parent might keep the over-saving behavior, e.g., because learning is slow for high values of consumption. Then, she would increase the inherited extreme wealth even further instead of dissaving it (the latter is what a rational agent would do). I find that this mechanism generates a lot of immobility in the economy. This is also consistent with the findings of \cite{charles2003correlation}, who find large wealth to be more persistent across generations.

To provide some connection with the data, I compare the results of the simulation with the estimates of \cite{chetty2014united} and \cite{charles2003correlation}. In \cite{chetty2014united}, the child is approximately 30 years old, and thus I use income after 30 periods for comparison. In  \cite{charles2003correlation}, the average age of children is 37.5 years, and thus I use wealth after 38 periods. 
However, the comparison with these empirical estimates is only suggestive. \cite{charles2003correlation} consider the wealth of the child before bequests, and in \cite{chetty2014united} bequests are also unlikely because of the young age. The Aiyagari framework does not allow for a good counterpart of their estimates, as it essentially assumes bequests.\footnote{That said, bequests in the real world might have limited importance for most agents except for the richest ones. It has also been documented that the latter tend to start transferring wealth earlier.} At the same time, all channels of intergenerational immobility other than wealth are not represented well in the model. For example, there is no transmission of human capital as productivity shocks have a relatively transitory effect on income. Thus, a direct comparison of estimates from the model and the papers might be impossible.

Table~\ref{table:mobility} presents the results. For the baseline NN economy, the mobility is  much lower, just as Figure~\ref{fig:chetty_2nd_gen} suggests. The moments generated by the model better fit the correlation between parental and child incomes than the RE baseline. Most of the difference in average incomes comes from the difference in interest income. Accordingly, the model also predicts a smaller mobility of wealth. The predictions are too small compared to the estimates of  \cite{charles2003correlation} because of the large role of bequests in the model.

 \begin{table}[h]
\begin{center}
\begin{tabular}{llcccccc}
\hhline{~~======}
&  &  Data  & Low rationality  & High rationality & RE \\ \hline
Income & &       &           &            &              \\
 &  Rank-rank slope & 0.341         &      0.370            &    0.132        &      0.125        \\
 &  Intergenerational elasticity &  0.45        &          0.35          &    0.13       &      0.13        \\
Wealth &    &     &           &            &              \\
 &  Shorrocks index  &  0.88        &      0.56             &    0.89       &      0.81      \\
 &  Intergenerational elasticity  &  0.37        &      0.49          &       0.32    &   0.46        \\ \hline \hline
\end{tabular}
\end{center}
\caption{Intergenerational mobility for NN agents with NN parents.} \medskip
\small 
Income is defined as $r a_t + w z_t.$ For income, the gap between parents and children is 10 years. The empirical estimates are from \cite{chetty2014united}. For wealth, the gap between parents and children is 18 years. The empirical estimates are from \cite{charles2003correlation}.
The Shorrocks index is defined as $\frac{n - \sum_{i=1}^n a_{i,i}}{n-1},$ where $(a_{i,j})$ is the transition matrix between quintiles. A smaller index indicates lower mobility. To find intergenerational elasticity of wealth, I exclude agents with no savings. 
\label{table:mobility}
\end{table}

Interestingly, for agents with high rationality I observe mobility even slightly higher than that in the RE economy. The mechanisms discussed above, e.g., the role of mistakes as an equalizing noise, outweigh the channel of over-saving by the children of extremely wealthy agents (who are almost absent among NN agents with high rationality; see the distribution in Figure~\ref{dist_high}), and learning again becomes an intergenerational equalizer.

\section{Conclusion}  \label{sec:conc}
Neural networks provide a new approach to model behavioral agents in macroeconomics. Its credibility comes from the fact that neural networks can successfully perform a large variety of human-like tasks. I focus on a particular advantage of neural networks: they can be used to model dynamic learning of a decision rule. This is a kind of learning that is not traditionally considered in economics, but it might be important as agents face it often in the real world. 
For the consumption-saving problem, the approach reveals a connection between the wealth experience and the quality of choices made. It also reveals learning traps in which agents become permanently poor (in terms of savings) or suboptimally rich. That might explain several patterns observed in the wealth distribution. 
This model can be scaled to multiple assets and more general economic control problems, potentially revealing some learning patterns and traps there too.

\bibliographystyle{jfe}
\bibliography{bibl}
\clearpage

\appendix

\section{Neural Networks}
\label{sec:nnets}
\begin{figure}[!htb]
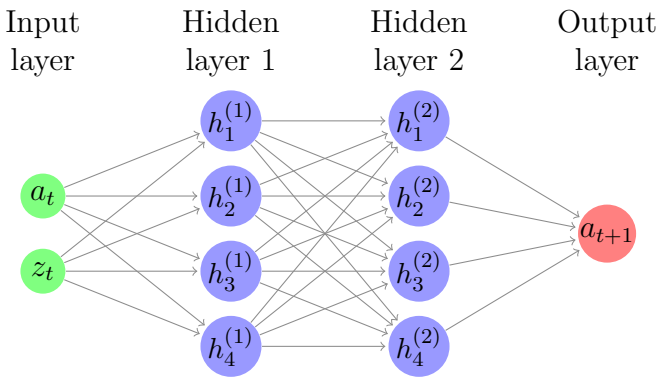

\begin{center}
\begin{neuralnetwork}[height=4]
        \newcommand{\x}[2]{\IfEqCase{#2}{%
        {1}{$a_t$}
        {2}{$z_t$}}}
        \newcommand{\y}[2]{$a_{t+1}$}
        \newcommand{\hfirst}[2]{\small $h^{(1)}_#2$}
        \newcommand{\hsecond}[2]{\small $h^{(2)}_#2$}
        \inputlayer[count=2, bias=false, title=Input\\layer, text=\x]
        \hiddenlayer[count=4, bias=false, title=Hidden\\layer 1, text=\hfirst] \linklayers
        \hiddenlayer[count=4, bias=false, title=Hidden\\layer 2, text=\hsecond] \linklayers
        \outputlayer[count=1, title=Output\\layer, text=\y] \linklayers
    \end{neuralnetwork}
 \end{center}
   \caption{An example of a neural network with two hidden layers.} \label{fig:nn}
 \end{figure}

This section is supposed to briefly introduce the notion of a neural network but cannot serve as a complete introduction to the subject.  An (artificial) neural network is a result of stacking several layers of neurons. Each neuron has an output, which is given by applying an activation function $f$ to a linear combination of the outputs from the neurons in the previous layer (for the first hidden layer, a linear combination of the original inputs). Figure~\ref{fig:nn} provides an example of a neural network relevant for this paper. This network has a two-dimensional input and a one-dimensional output. It has two hidden layers with $n_1 = n_2 = 4$ neurons in each of them. Formally, the system is given by

$$h_{j}^{(1)} = f\big(\theta_{j,0}^{(1)}+ \theta_{j,1}^{(1)}a_t + \theta_{j,2}^{(1)}z_t \big),$$$$ h_{j}^{(2)} = f\big(\theta_{j,0}^{(2)} + \sum_{i=1}^{n_1} \theta_{j,i}^{(2) } h^{(1)}_i\big), $$$$a_{t+1} = \theta_{0}^{(3)} + \sum_{i=1}^{n_2} \theta_{i}^{(3) } h^{(2)}_i.$$

 An activation function $f$ is needed to make the system non-linear, otherwise, the output would be simply a linear function of the original inputs. These days, the standard choice for $f$ is the ReLU activation function, which is given by $f(x) = \max \{0,x\}$. It also has a biological motivation, see \cite{hahnloser2000digital}.

There are many reasons why neural networks became extremely successful in practical applications. Some of these reasons are yet to be understood from the theoretical viewpoint, but some are already established. \cite{rumelhart1986learning} introduced back-propagation, the procedure for efficient differentiation with respect to the weights of a network. \cite{cybenko1989approximation}, \cite{barron1993universal}, \cite{bach2017breaking}, and others emphasize the good approximation properties. The modern strand of the literature, including \cite{ba2013deep}, emphasizes the better ability of deep neural networks to \textit{learn} while having expressivity similar to that of networks with just one hidden layer.
 
There is a feature of neural networks that is related to the described learning process. Neural networks are models with many parameters and high expressivity. In terms of the bias-variance trade-off, this means that they usually have low bias but a very high variance. For example, a typical neural network used for a classification task could perfectly fit the data set it is trained on. To avoid overfitting, training is stopped after some number of gradient descent steps. Similarly, neural networks behind saving policy functions are updated incrementally. The same approach is implemented in all modern deep reinforcement learning algorithms.

\section{Asymptotic Rationality}
\label{sec:app_rationality}

In the following, I provide evidence that the suggested model of learning is asymptotically rational. Though no formal statement about convergence is provided, I show that agents with hyperparameters that reflect few cognitive limitations can learn the optimal policy with a very high degree of precision. 
This suggests that the learning described in this paper is indeed directed toward the optimal policy, and the class of neural network agents nests very good approximations to rational agents. If this property was not satisfied, that would cast doubt on the credibility of the model, as too many updating schemes satisfy the other properties listed on page~\pageref{learning_properties}.
\begin{figure}[t]
\begin{center} \includegraphics[width=12cm]{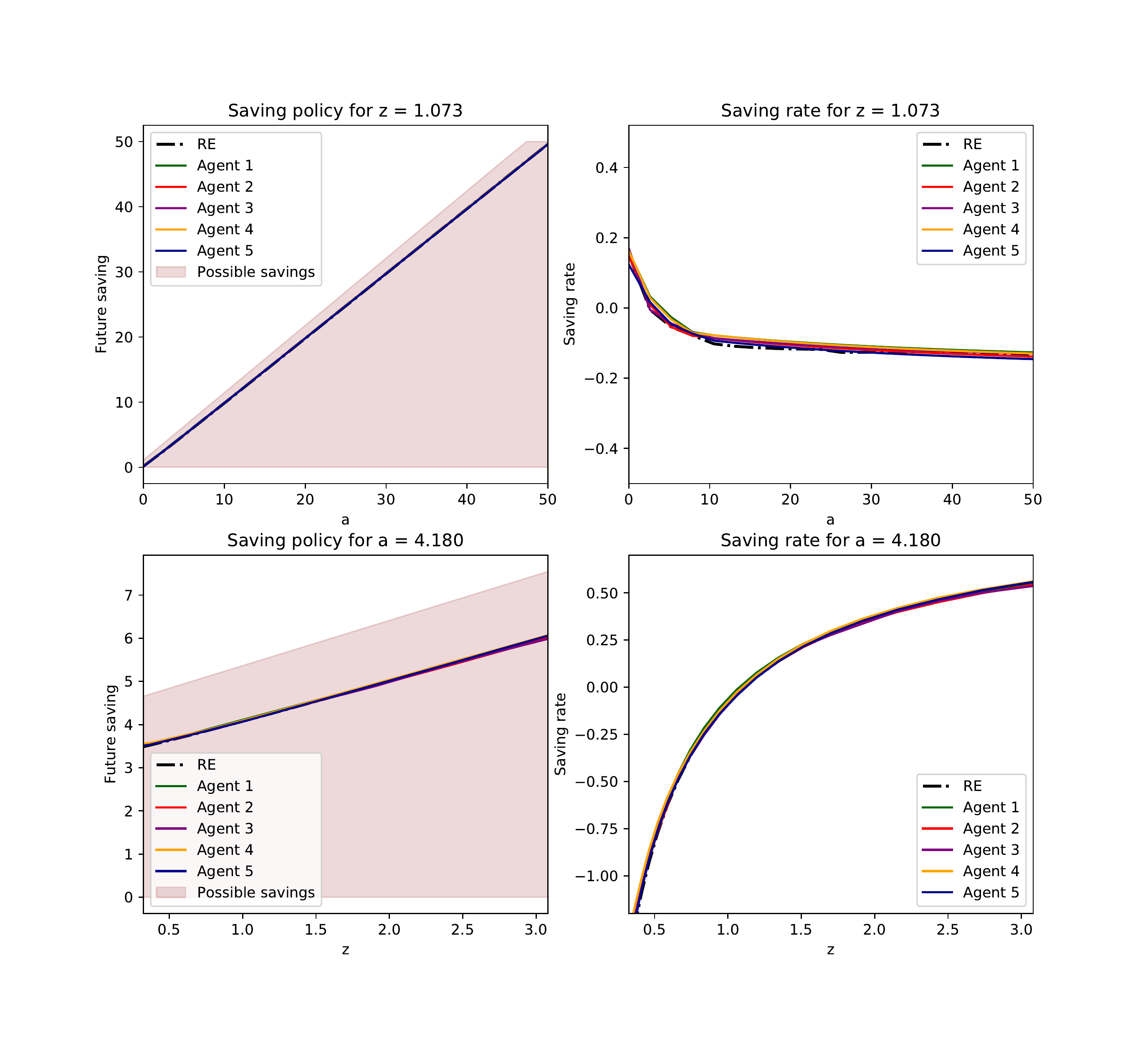}
\caption{Examples of successful learning.} \medskip
\small In the first row, $z$ is fixed at the mean level. In the second row, $a$ is fixed at the mean level for the RE economy. Saving rate is defined as $\frac{a_{t+1}-a_t}{ra_t + wz_t}.$ 
\label{fig:asymp_rat}
\end{center} \end{figure}

For this experiment, I use a more expressive network architecture with 4 hidden layers that have 4, 8, 8, and 4 neurons respectively. I also use a smaller learning rate $\alpha = 0.001$. I simulate several agents who learn from external experiences only (values of $a_t$ are distributed uniformly over the savings dimension of the state space) and do that at a high frequency, making 10,000 learning steps per period.  Figure~\ref{fig:asymp_rat} illustrates policies of the first 5 agents at the end of their lives. 
As can be seen, the degree of precision is very high even when policies are plotted as saving rates, let alone plain values of savings.


\end{document}